\documentclass{aa}  
\usepackage{graphicx}
\usepackage{txfonts}
\usepackage{graphicx}	
\usepackage{amsmath}	
\usepackage{multirow}
\usepackage{threeparttable}
\usepackage[T1]{fontenc}
\usepackage{hyperref}
\hypersetup{
    colorlinks = true,
    urlcolor  = blue,
    linkcolor = blue,
    citecolor = blue
    }

\usepackage{orcidlink}
\let\orcid\orcidlink

\begin{document} 
   \title{Jet reorientation revealed by intermittent jet activity in radio galaxy 0954+556}
   \author{Ai-Ling Zeng \orcid{0009-0000-9427-4608}
    \inst{\ref{SHAO},\ref{ShanghaiTech},\ref{iaa}}$^{,\dagger}$
          \and Wei Zhao \orcid{0000-0002-1992-5260}
          \inst{\ref{SHAO},\ref{ucas}}$^{,\dagger}$ 
          \and Jun Yang \orcid{0000-0002-2322-5232}
          \inst{\ref{Chalmers}}
          \and Xu-Zhi Hu \orcid{0000-0002-5398-1303}
          \inst{\ref{yunnan}}
          \and Furen Deng \orcid{0000-0001-8075-0909}
          \inst{\ref{ucas}, \ref{nao}}
          \and Yu Lei \orcid{}
          \inst{\ref{SHAO},\ref{ucas}}
          \and Xiaoyu Hong \orcid{0000-0002-1992-5260}
          \inst{\ref{SHAO},\ref{ShanghaiTech},\ref{ucas}}$^{,\dagger}$
          \and Xiang Liu \orcid{0000-0001-9815-2579}
          \inst{\ref{ucas},\ref{xao}}
          \and Liang Chen \orcid{0000-0002-1908-0536}
          \inst{\ref{SHAO},\ref{ucas}}
          \and Mai Liao \orcid{0000-0002-9137-7019}
          \inst{\ref{nao},\ref{cas-sac},\ref{chile}}
          \and Xiaolong Yang \orcid{0000-0002-4439-5580}
          \inst{\ref{SHAO},\ref{ucas}}
          \and Hai-Tian Shang \orcid{0009-0002-8909-2935} 
          \inst{\ref{SHAO},\ref{ShanghaiTech}}
          }
\institute{
Shanghai Astronomical Observatory, Chinese Academy of Sciences, Shanghai 200030, China\label{SHAO} $^{\dagger}$\email{azeng@iaa.es}     
\and School of Physical Science and Technology, ShanghaiTech University, Shanghai 201210, China\label{ShanghaiTech} 
\and Instituto de Astrof\'{i}sica de Andaluc\'{i}a-CSIC, Glorieta de la Astronom\'{i}a s/n, E-18008 Granada, Spain\label{iaa} 
\and School of Astronomy and Space Science, University of Chinese Academy of Sciences, Beijing 100049, China \label{ucas}
\and Department of Space, Earth and Environment, Chalmers University of Technology, Onsala Space Observatory, SE-439\,92 Onsala, Sweden \label{Chalmers}
\and School of Physics and Astronomy, Yunnan University, Kunming 650500, China \label{yunnan}
\and National Astronomical Observatories, Chinese Academy of Sciences, Beijing 100101, China \label{nao} 
\and Xinjiang Astronomical Observatory, Chinese Academy of Sciences, 150 Science-1 Street, Urumqi 830011, China \label{xao}
\and Chinese Academy of Sciences South America Center for Astronomy, National Astronomical Observatories, CAS, Beijing, 100101, China \label{cas-sac}
\and Instituto de Estudios Astrofísicos Facultad de Ingeniería y Ciencias Universidad Diego Portales Av. Ejército 441, Santiago, Chile \label{chile}
}

   \date{Received December 29, 2024; accepted October 17, 2025}
 
  \abstract
   {Intermittent jet activity of active galactic nuclei (AGNs) is a common phenomenon, whereas significant jet reorientation during episodic jet activity in relatively young radio galaxies are rarely reported. The quasar 0954+556 at z=0.903 is an intriguing source exhibiting an unusual radio jet structure with significantly different jet directions at kiloparsec (kpc) and parsec (pc) scales. At kpc scales, images from the Very Large Array (VLA) exhibit a bright core, a linear jet extending $\sim$24 kpc to the northwest, and a discrete jet component $\sim$16 kpc to the northeast. At pc scales, images from the Very Long Baseline Array (VLBA) show a two-component structure with a projected separation of $\sim$360 pc in the north--south direction. }
   {The peculiar structure of 0954+556 might result from jet reorientation. Here, our aim was to investigate the possible mechanism via multiscale and multifrequency deep radio images.}
   {We performed VLA and VLBA observations of 0954+556. Together with some existing data in the NRAO data archive, we made multiple VLA images at 1.4--22~GHz and VLBA images at 1.7--43 GHz for various image analyses of the jet structure.  } 
   {We identified the location of the radio core at pc scales, detected the faint counter-jets at both pc and kpc scales for the first time, and revealed a diffuse emission region connecting pc- and kpc-scale forward jets. Our spectral index distribution and spectral aging analysis indicate that 0954+556 might undergo at least two episodes of jet activity during the current AGN phase. Moreover, pc-scale polarization maps display a well-resolved spine-sheath polarization structure.  } 
   {It seems that the jet direction of 0954+556 changed significantly during intermittent jet activity. This may explain the different jet orientations and spectral ages observed from kpc to pc scales. The research provides a strong case that AGN jet direction might change rapidly on timescales of one million years.}


   \maketitle

\section{Introduction}
\label{sec:intro}
It is widely accepted that nearly all galaxies in the Universe host supermassive black holes (SMBHs) at their centers \citep{1998Natur.395A..14R}. Most galaxies are dormant due to the lack of fuel-feeding central black holes. Active galactic nuclei \citep[AGNs,][]{1995PASP..107..803U} are a small fraction that produce high luminosity in a concentrated region at the center of a galaxy where the SMBH is accreting significant amounts of gas. As a result of accretion, the magnetic field anchored in the rotating disk drives winds that carry energy and angular momentum away from the disk, leading to jet formation and outflows \citep[BP model,][]{1982MNRAS.199..883B}. Alternatively, the energy and angular momentum are extracted from a spinning black hole by a purely electromagnetic mechanism \citep[BZ model,][]{1977MNRAS.179..433B}.

It was a question of whether the active state of a galaxy is a repeated phenomenon or a one-time and long-term activity, after which the galaxy never reactivates.
\citet{2001ApJ...547...90H} studied all galaxies, and found they go through different phases of nuclear activity over their lifetimes, which can be supported by the observational fact that there are more young sources than sources with extended and old structures \citep[see Section 1 in ][]{2009ApJ...698..840C}. 
The recurrent activity can be recognized when new activity occurs and the emissions of past episodes remain for a long time.
Consequently, a young radio source embedded in a relic structure is expected.

Intermittent jet activity has been reported many times. Double-double radio galaxies \citep[DDRG,][]{2000MNRAS.315..371S} and X-shaped radio galaxies \citep[XRG,][]{2007AJ....133.2097C} serve as prime examples of such intermittent phenomena. A DDRG consists of a pair of double lobes with a common center, the outer lobes being the relic of past activity and the inner lobes being a manifestation of new jet activity. 
To date, four cases of triple-double radio galaxies, which contain three pairs of double lobes resulting from three separate episodes, have been identified: B0925+420 \citep{2007MNRAS.382.1019B}, J1409-0302 \citep{2011MNRAS.417L..36H}, J1216+0709 \citep{2016ApJ...826..132S}, and J1225+4011 \citep{2023MNRAS.525L..87C}. An XRG displays a conventional pair of active lobes and an additional pair of low surface brightness "wings" that are considered as remnants from a rapid realignment.
In addition to these two types, signs of intermittent activity may be present in young radio galaxies that are in the nascent stage of radio galaxy evolution, reflecting the initial phases of their radio emission. 

In addition to the observational evidence of intermittent jet activity, many theoretical studies have proposed possible explanations for AGNs switching on and off. 
Mergers \citep{2002Sci...297.1310M,2003ApJ...594L.103G,2008ASPC..386..467B} 
may play a significant role on long timescales ($10^6$--$10^8$ years). 
The infalling galaxy passes through the host galaxy and turns around every $\sim$$10^7$ yr, causing instabilities in the host galaxy during each pass. After several encounters, the colliding galaxies merge completely, usually taking up $\sim$$10^8$ yr.
Ionization instabilities in a narrow and unstable zone of the accretion disk can propagate throughout the disk, resulting in intermittent activity on timescales of $10^6$--$10^8$ yr \citep{1996ApJ...458..491S}. Additionally, the short timescale activity is related to radiation pressure instabilities in the disk and mainly operates on timescales $10^4$--$10^6$ yr \citep{2002ApJ...576..908J,2009ApJ...698..840C,2017A&A...603A.110G}.

For this work we focused on 0954+556 (also named 4C +55.17, $z$= 0.903,\footnote{\url{https://www.cv.nrao.edu/MOJAVE/sourcepages/0954+556.shtml}} \citealt{2020ApJS..249....3A}), which was suggested as a case of intermittent activity due to its peculiar morphology \citep{2005A&A...434..449R}. 
0954+556 has been categorized as a flat-spectrum radio quasar (FSRQ) because of the presence of broad optical emission lines in its spectrum, high optical$/$UV core luminosity (absolute $B$-band magnitude, $M_B <-23$), and high $\gamma$-ray luminosity ($L_{\gamma}\simeq 10^{47}$ erg s$^{-1}$) \citep{1995ApJ...447..139W, 2006A&A...455..773V, 2008ApJS..175..297A, 2011ApJ...738..148M}.
At parsec (pc) scales, two components separated by $\sim$50 mas were detected by the Very Long Baseline Array (VLBA) at 1.7 and 4.8 GHz, resembling a double oriented in the north--south direction. The total extent corresponds to a projected linear size of 640\,pc (80\,mas). In contrast, at kiloparsec (kpc) scales, 0954+556 presents a triple structure with a projected linear size up to 36\,kpc, i.e., 4.5\,arcsec, under the detection of the Very Large Array (VLA) at 1.4\,GHz. At 5\,GHz, the triple structure is resolved into a bright core, a linear jet extending northwest for approximately 3 arcsec, and a discrete component 2\,arcsec northeast of the core. As the parsec-scale structure coincides with the kiloparsec-scale core,
the morphology of 0954+556 has naturally raised a question about the reason for the jet changing direction between the two scales. 

The classification of 0954+556 is controversial. It meets the criteria of compact symmetric objects \citep[CSOs,][]{1996ApJ...460..612R,2024ApJ...961..240K}, featuring a very low radio variability and a symmetric double structure with a linear size less than 1 kpc in VLBA images, similar to the classical doubles of Fanaroff-Riley Class II. Accordingly, it has been classified as a case of a CSO at parsec scales, and it was considered to be a case of a medium symmetric object (MSO, the larger version of CSOs, with a linear size ranging from 1 to 20\,kpc) at kiloparsec scales, as the radio flux is dominated by the core region, i.e., the CSO part. \cite{2011ApJ...738..148M} further argued the preference of 0954+556 as a CSO. As a subclass of young radio galaxies, CSOs could be interpreted in several ways \citep{2021A&ARv..29....3O}: as sources in their early stages of evolution, as sources confined by a dense interstellar medium, or as sources experiencing intermittent jet activity. 
Therefore, whether this source arises from intermittent activity deserves further investigation.

Previous radio studies of 0954+556 focused primarily on total intensity maps, due to a lack of data, thus leaving the above questions unanswered. For this work we further investigated the source by analyzing the total intensity, spectral index distribution, spectral aging, and polarized emissions, using data from VLA and VLBA over a wide frequency range from 1.4 to 43\,GHz. From the perspective of CSOs, we explored the intermittent jet activity of 0954+556 and how the intermittent activity affects source morphology.

The paper is organized as follows. Section~\ref{sec:obsdr} summarizes the observations and data reduction. Section~\ref{sec:res} presents the obtained multifrequency images and results of analysis on variability at $\gamma$-ray and radio wavelengths. In Section~\ref{sec:spa} the results of the spectral analysis are presented. We also present the results of VLBI polarimetry at 1.7, 5.0, and 8.4\,GHz in Section\,\ref{sec:pol}. Section~\ref{sec:dis} includes the discussion and Section~\ref{sec:sum} gives a summary.
Cosmological parameters are adopted as follows: $H_0=67.4 \mathrm{~km} \mathrm{~s}^{-1} \mathrm{Mpc}^{-1}$, $\Omega_{\mathrm{M}}=0.315$, $\Omega_{\Lambda}=0.685$ \citep{2020A&A...641A...6P}, and therefore 1\,mas $\sim$ 8\,pc. Throughout the paper, $S_{\nu}\propto\nu^{-\alpha}$, in which $S_{\nu}$ represents the flux density, $\nu$ is frequency, and $\alpha$ is the spectral index.

\section{Observation and data reduction}
\label{sec:obsdr}
\subsection{Observations}
\label{ssec:obs}
Over the last two decades, we have observed 0954+556 multiple times either as the target source or as the phase-reference calibrator,
thus a great deal of VLA and VLBA data have been accumulated by now, including the following: two VLA datasets at 8.4 and 22.5\,GHz from December 2000 (observation code: AH0721); one VLBA dataset at 1.7\,GHz from February 2000 (BH065), two VLBA datasets at 5 and 8.4 GHz acquired in July 2002 (BH096), all designed for polarimetry; and three VLBA datasets at 5, 22, and 43\,GHz in mid-2016 (BZ061, BL222). 
Furthermore, we obtained two additional VLA datasets at 1.4 and 5\,GHz in November 2000 (AM0672), one VLA dataset at 15\,GHz in October 1996 (TESTT), as well as five VLBA datasets 
at 5\,GHz between September 2002 and August 2005 (BM170, BM208) from the NRAO archive.\footnote{\url{https://archive.nrao.edu/archive/advquery.jsp}} Table~\ref{tab:img} lists the information of each observation, such as frequency, observation code, array, and observation date.

To make a brief analysis of the variability of 0954+556, we collected $\gamma$-ray photon flux data from the Fermi-LAT Light Curve Repository (LCR)\footnote{\url{https://fermi.gsfc.nasa.gov/ssc/data/access/lat/LightCurveRepository/source.html?source_name=4FGL_J0957.6+5523}} \citep{2023ApJS..265...31A} and the total flux at radio band from the Owens Valley Radio Observatory (OVRO) 40m Telescope Monitoring Program \citep{2014MNRAS.438.3058R}. 
The $\gamma$-ray data were divided into seven-day bins, covering a period between
8 August 2008 and 6 December 2024; the data represent the average integrated
photon flux with energies from 0.1 to 100 GeV.
In the OVRO program 0954+556 was observed at 15\,GHz every four days from 19 April 2009 to 30 December 2011.

\subsection{Data reduction}
\label{ssec:dr}

The data were calibrated using the NRAO Astronomical Image Processing System (\texttt{AIPS}) \citep{2003ASSL..285..109G} 
following the standard procedures described in the \texttt{AIPS} cookbook.\footnote{\url{http://www.aips.nrao.edu/cook.html}} The data from VLA were inspected and edited using \texttt{LISTR}, \texttt{QUACK}, and \texttt{UVFLG}, followed by flux and phase calibration. The flux calibrator was set to 3C\,48 or 3C\,286 in our calibration by \texttt{SETJY} and \texttt{CALRD}. Finally, the task \texttt{CALIB} was used to calibrate amplitudes and phases.

Following Appendix C in the \texttt{AIPS} cookbook, we calibrated data from VLBA in \texttt{AIPS}. On each track of VLBA data at 1.7, 5, and 8.4\,GHz, we performed global fringe fitting twice for 0954+556 to correct the structure-induced phase errors, considering the pc-scale structure of 0954+556 at these frequencies is extended. 
For the first global fringe fitting, we solved the residual delay, rate, and phase by assuming a point source model. 
The output $uv$ data was transformed into images with a clear structure of 0954+556 by the iteration of CLEAN and phase self-calibration in \texttt{Difmap} \citep{1997ASPC..125...77S}.
The second global fringe fitting was performed by using the cleaned images.
Then we applied the resulting solutions containing corrections for the structure-induced phase errors to 0954+556. 
Once the initial calibration was completed in \texttt{AIPS}, self-calibration (both phase and amplitude) and imaging loops 
were performed using \texttt{Difmap} to obtain images with high dynamic ranges. 

The 22\,GHz VLBA data were imaged twice after global fringe-fitting.
First, we imaged the data in the full $uv$ range (2.3-636 $\mathrm{M}\lambda$),
and only diffuse emissions were detected marginally.
The ratio of peak flux density to the noise level in the cleaned image was so low ($\sim$6) 
that it was infeasible to perform self-calibration.
In the second trial, the maximum $uv$ range was limited to 48.0\,$\mathrm{M}\lambda$ (see detailed description in Section \ref{sssec:prsp}).
On the resultant image, the peak-to-noise ratio was raised to 20. 
Phase-only self-calibration was applied very carefully for several iterations to recover the flux density. 

At 43\,GHz, in global fringe fitting, solutions with a signal-to-noise ratio (S/N) higher than 5 (the default S/N threshold in the \texttt{AIPS} task \texttt{FRING}) only took 10\%, so we adjusted the threshold of S/N to 4, and then the good solutions increased to 20\%.
The good solutions were applied to the data, and a tentative 43\,GHz VLBI image was produced.
As the peak-to-noise ratio of the resultant image was only 10, no self-calibration was performed.

The resulting images are presented in Figs. \ref{fig:vla_figure} and \ref{fig:vlba_figure}. The parameters of the images (e.g., frequency, array, date, $uv$ range, size of the synthesized beam, rms noise level in a clean image, peak intensity, and total clean flux density) are listed in Table~\ref{tab:img}.

\subsection{Polarization calibrations}
\label{ssec:polc}
We performed polarization calibration in \texttt{AIPS} on VLBA observations designed for polarimetry, i.e., 1.7\,GHz data obtained in February 2000 (BH065) and 5\,GHz and 8.4\,GHz data obtained in July 2002 (BH096).
First, we used the \texttt{AIPS} task \texttt{RLDLY} to solve the R-L delays (right circular and left circular polarization).
The instrumental polarization (D-terms) was solved by running the \texttt{AIPS} task \texttt{LPCAL}
on a weakly polarized source, which is DA\,193 in project BH065, and OQ\,208 in project BH096. 

The electric vector position angle (EVPA) of the linear polarization was calibrated by comparing the integrated EVPA of a calibrator obtained in our observation with the reference EVPA of the calibrator reported by the Very Large Array polarization monitoring program\footnote{\url{http://www.vla.nrao.edu/astro/calib/polar/}} in nearby epochs. 
For the project BH096 observed on 20 July 2002, J2253+1608 (3C\,454.3) was used as the calibrator for EVPA calibration at 5 and 8.4\,GHz \citep{2024ApJ...965...74H}. The nearest epoch of J2253+1608 in the Very Large Array polarization monitoring
program was on 10 July 2002, and the differences between the reference EVPA and the integrated EVPA of J2253+1608 were $29.8^{\circ}\pm2.6^{\circ}$ at 5\,GHz and $-41.9^{\circ}\pm2.0^{\circ}$ at 8.4\,GHz.

As explained in detail by \citet{2024ApJ...974..111H}, there is no information close to the observational date of the project BH065, so it is impossible to recover the absolute EVPA distribution of 0954+556 at 1.7\,GHz by using an EVPA calibrator. While the absolute EVPAs at 1.7 GHz are
arbitrary, the relative EVPAs across the entire total intensity structure are accurate. Our results reveal the EVPA distributions of 0954+556 at 5 and 8.4 GHz are consistent (Section~\ref{sec:pol}), indicating that the rotation measure (RM) is close to zero. In addition, at kpc scales, the RM of 0954+556 was estimated to be $+1\pm1\,\mathrm{rad}\,\mathrm{m}^{-2}$ in \citet{1981ApJS...45...97S}. Assuming that the polarization remained stable from 2000 to 2002, and given the RM value, we can recover an approximate absolute EVPA distribution at 1.7\,GHz by aligning it with those at higher-frequency maps. Specifically, we rotated the EVPAs at 1.7\,GHz to match the orientation at 5\,GHz, using the southern lobe as a reference. The resulting EVPA distributions at three frequencies are presented in Section~\ref{sec:pol}, and the image at 1.7\,GHz generally represents the absolute EVPA distribution at this frequency.

\subsection{Preparations for spectral analysis}
\label{sssec:prsp}

The spectral index distribution map requires the alignment of images. The core of an extragalactic jet is a region with an optical depth $\tau_s \approx 1$. Due to synchrotron self-absorption, the core position varies with the observing frequency and follows the relation $r_{\text{core}} \propto \nu^{-1/k_r}$ \citep{1979ApJ...232...34B}, which is a phenomenon known as the core-shift effect \citep{1979ApJ...232...34B,2008A&A...483..759K,2011A&A...532A..38S}. In contrast, the jet part is typically optically thin, showing a steep spectrum. As the optically thin jets are assumed not to change with frequency, it is used to align images at different frequencies. Aligning optically thin compact components across images is the most common way to correct for the core shift \citep{2014AJ....147..143H}. In terms of sources lacking compact features and showing diffuse structures, alignment is instead performed using the two-dimensional cross-correlation method \citep{2008MNRAS.386..619C}, which adopts spatial correlations of optically thin extended jet structure to register images.

To avoid errors brought by different $uv$ samplings and resolutions, 
the images used in spectral index maps and the spectral aging analysis were produced with data in the same $uv$ range, in which the minimum $uv$ distance is the smallest value of the highest frequency and
the maximum is the largest value of the lowest frequency. In addition, images were set to the same map size
and cell size, and restored with a beam of the lowest frequency. Pixels with flux density below 5.4$\sigma_{rms}$, alongside those with spectral index uncertainty exceeding 40\%, were blanked when we produced the spectral index maps.

To produce the kpc-scale spectral index map, the 1.4 and 5 GHz VLA data were restricted to a common $uv$ range of $5.1-168.7 \mathrm{k}\lambda$ and were inverse Fourier transformed into images with a beam of $1449\times1941\,\mathrm{mas}, -85.13\degr$ to obtain the 1.4--5 GHz spectral index map.
For the 5 and 8.4\,GHz VLA data in a $uv$ range of $11.3-561.2 \mathrm{k}\lambda$, the inverse Fourier transform was used, and the resultant images were restored with a beam of $445.3\times587.2\,\mathrm{mas}, -81.7\degr$.

To produce pc-scale spectral index maps, 
1.7, 5 (20 July 2002), and 8.4\,GHz VLBA data were used.
For the data in a $uv$ range of $2.3-48.0 \mathrm{M}\lambda$, the inverse Fourier transform was used and the images were restored with a beam of $5.4\times7.3\,\mathrm{mas}, -15.5\degr$ to produce 1.7--5\,GHz index map;
for data in a $uv$ range of $5.4-39.5 \mathrm{M}\lambda$, the inverse Fourier transform was used and the images were restored with a beam of $4.3\times7.0\,\mathrm{mas}, -26.0\degr$ to produce 5--8.4\,GHz index map.

To produce kpc-scale images for spectral fitting in Section~\ref{ssec:ra}, VLA data in a $uv$ range of $30.8-168.6 \mathrm{k} \lambda$ were utilized, and the resulting images were restored with a beam of $1257\times1676\,\mathrm{mas}, -84.4\degr$.
At pc scales, VLBA data at 1.7, 5, 8.4, 22\,GHz incorporating the 15\,GHz VLBA data from \citet{2005A&A...434..449R} 
in a $uv$ range of $7.1-48.0 \mathrm{M}\lambda$ were utilized, and the resulting images were restored with a beam of $5.0\times6.5\,\mathrm{mas}, -18.3\degr$. 
As listed in Table \ref{tab:img}, the $uv$ range of data at 43\,GHz is $22.3-1105.4 \mathrm{M}\lambda$. If the 43\,GHz data is added into spectral analysis, the $uv$ range used in the spectral analysis will be $22.3-48.0 \mathrm{M}\lambda$. Many visibilities in every dataset will be flagged. Thus, data at 43\,GHz was not used in the spectral analysis.

\begin{table*}
    \caption{Image parameters}
    \label{tab:img}
    \begin{threeparttable}
    \begin{tabular*}{0.99\linewidth}{ccccccccccc} 
    \hline 
    Freq & Obs. Code& Array & Date  & $uv$ range &  \multicolumn{3}{c}{Synthesized Beam Size }&$\sigma_{\mathrm{rms}}$ & $S_{\text{peak}}$ & $S_{\text{int}}$  \\ 
         & &       &      &    & $B_{\mathrm{maj}}$ &$B_{\mathrm{min}}$ &$B_{\mathrm{pa}}$ &                        &                   &   \\
    (GHz) & &      &(dd/mm/yy) & &(mas)               &(mas)              &$\left({ }^{\circ}\right)$ &(mJy/beam)      &(Jy/beam)          &(Jy) \\ 
    (1)&(2)&(3)&(4)&(5)&(6)&(7)&(8)&(9) &(10)&(11)\\
    \hline
    1.4 &AM0672& VLA & 05/11/2000 & 1.5-168.7$k\lambda$ & 1959 &1462 & -85.1 & 0.7& 2.52 &2.99 \\
    5&AM0672& VLA & 05/11/2000 &5.1-561.2$k\lambda$ & 589 &447 & -81.7& 0.2& 1.80&2.05 \\
    8.4 &AH0721$^{*}$ & VLA & 12/01/2000 &11.3-922.5$k\lambda$&  485 &241 & 75.8& 0.2& 1.30&1.49 \\
    15 & TESTT& VLA & 12/10/1996 &14.0-1790.5$k\lambda$ & 153 & 133 & 25.9 & 0.4 & 0.84 &0.99 \\ 
    22 &AH0721$^{*}$& VLA & 12/01/2000 &30.8-2452.0$k\lambda$ & 175 &93.5 &78.8 &0.3 &0.73 &0.88 \\
    \hline
    1.7 &BH065$^{*}$& VLBA & 07/02/2000 &0.9-48.0$M\lambda$ & 7.51 &5.60 & -15.6 &0.3 &0.29 &2.03 \\ 
    5 &BZ061$^{*}$& VLBA & 13/06/2016 &1.7-52.4$M\lambda$ & 5.41 &4.89 &61.1 &0.2 &0.16 &1.14\\ 
    8.4 &BH096$^{*}$& VLBA & 20/07/2002 &5.5-159.1$M\lambda$ & 4.59 &2.88 &-3.2 &0.2 &0.13 &0.57\\ 
    22 &BZ061$^{*}$& VLBA & 09/07/2016 &2.3-48.0$M\lambda$ & 3.42 & 2.71 & -60.8 & 0.9 & 0.11 &0.29 \\
    43 &BL222$^{*}$& VLBA & 21/06/2016 &22.3-1105.4$M\lambda$ & 0.47 & 0.31 & 26.7 & 0.9 & 0.016  & 0.017 \\
    \hline 
    \end{tabular*}
\begin{tablenotes}[flushleft] 
    \footnotesize  
    \item \textbf{Notes.} The columns are as follows: (1) observation frequency; (2) observation code; the datasets marked with an asterisk represent our observations; (3) VLBI array used for observation; (4) observation epoch; (5) $uv$ range of visibilities in the observational data. Notably, the $uv$ range of the 22\,GHz VLBA data was cut to get a reliable image; (6)$\sim$(8) major axis of restoring beam, minor axis of restoring beam, and position angle of major axis, respectively; (9) rms noise level of an image; (10) peak intensity of an image; (11) total flux density of an image.
\end{tablenotes}
\end{threeparttable}
\end{table*}

 \begin{figure*}
 	\includegraphics[width=1.0\textwidth]{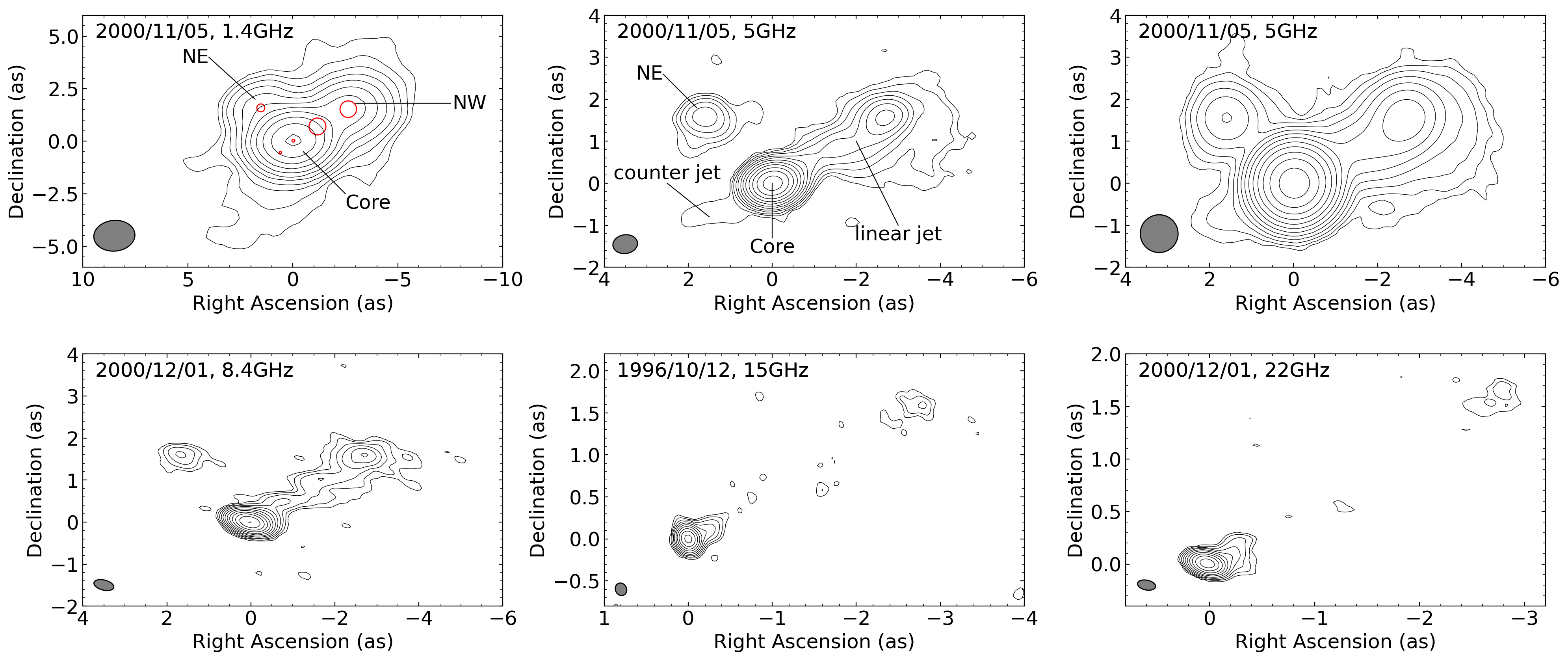}
     \caption{VLA flux density images of 0954+556 at kpc scale. The contours are drawn at -1, 1, 2, 4, 8, ..., of the first contour level (3$\sigma_\mathrm{rms}$). The synthesized beams are plotted in the bottom left corner of each image. In the top left panel, the red circles overlapping the contours present the results of modelfit. In the top middle panel, the 5\,GHz image shows a weak counter-jet. To better reveal the weak counter-jet, the 5 GHz image was restored with a circular beam of 900 mas, as shown in the top right panel.}
     \label{fig:vla_figure}
 \end{figure*}

 \begin{figure*}[h!]
 \centering
 	\includegraphics[width=0.8\textwidth]{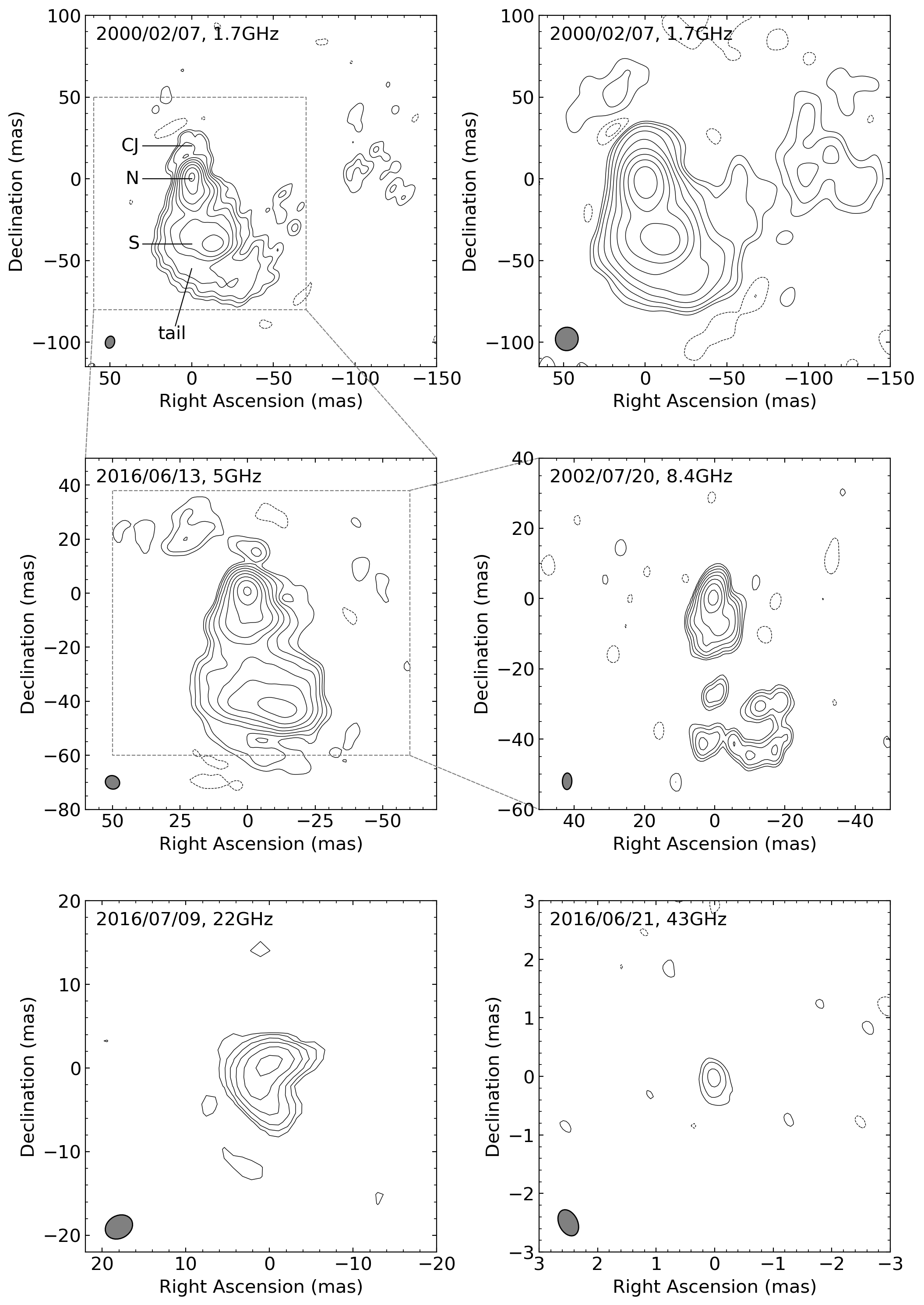}
     \caption{VLBA flux density images of 0954+556 at pc scale. The contours are drawn at -1, 1, 2, 4, 8, ..., of the first contour level (3$\sigma_\mathrm{rms}$). The synthesized beams are also plotted in the bottom left corner of each image. The image in the top right corner is the tapered image at 1.7\,GHz (\texttt{UVTAPER 0.5,\,10}). 
     }
    \label{fig:vlba_figure}
 \end{figure*}

 \begin{table}
    \centering
    \caption{Parameters of the fitted circular Gaussian models for the VLA image at 1.4\,GHz}
    \label{tab:modelfit}
    \begin{threeparttable}
    \begin{tabular*}{1.0\linewidth}{ccccc} 
    \hline 
    Comp &$S_{\mathrm{tot}}$ &$d$ &$r$ & $\theta$\\
     & (mJy)              & (mas)   &(mas)  & ($^{\circ}$)\\
    (1) & (2) & (3) & (4) & (5)   \\ 
    \hline
    NE          & 62$\pm$15    & 377$\pm$69   & 2200$\pm$35        & 43.9$\pm$0.9\\
                &28$\pm$3       & 107$\pm$8   & 800$\pm$4        & 132.0$\pm$0.3\\
    core        &2499$\pm$94    & 134.0$\pm$3   & 38$\pm$2     & -40.5$\pm$2.7\\
                &111$\pm$21   & 802.8$\pm$75   & 1365$\pm$37      & -59.0$\pm$1.6\\
    NW          &264$\pm$31    & 787.2$\pm$78   & 3048$\pm$39      & -60.0$\pm$0.7\\
    \hline 
    \end{tabular*}
    \begin{tablenotes}[flushleft] 
    \footnotesize  
    \item \textbf{Notes.} The columns are as follows: (1) name of each component; (2) total flux density of the fitted Gaussian in mJy; (3) FWHM major axis of the fitted Gaussian in mas; (4) distance of a Gaussian model from the phase center of an image in mas; (5) position angle of the fitted Gaussian from the phase center in degrees (positive values for north by east).
    \end{tablenotes}
    \end{threeparttable}
\end{table}

 \begin{figure*}
 \centering
 	\includegraphics[width=0.7\textwidth]{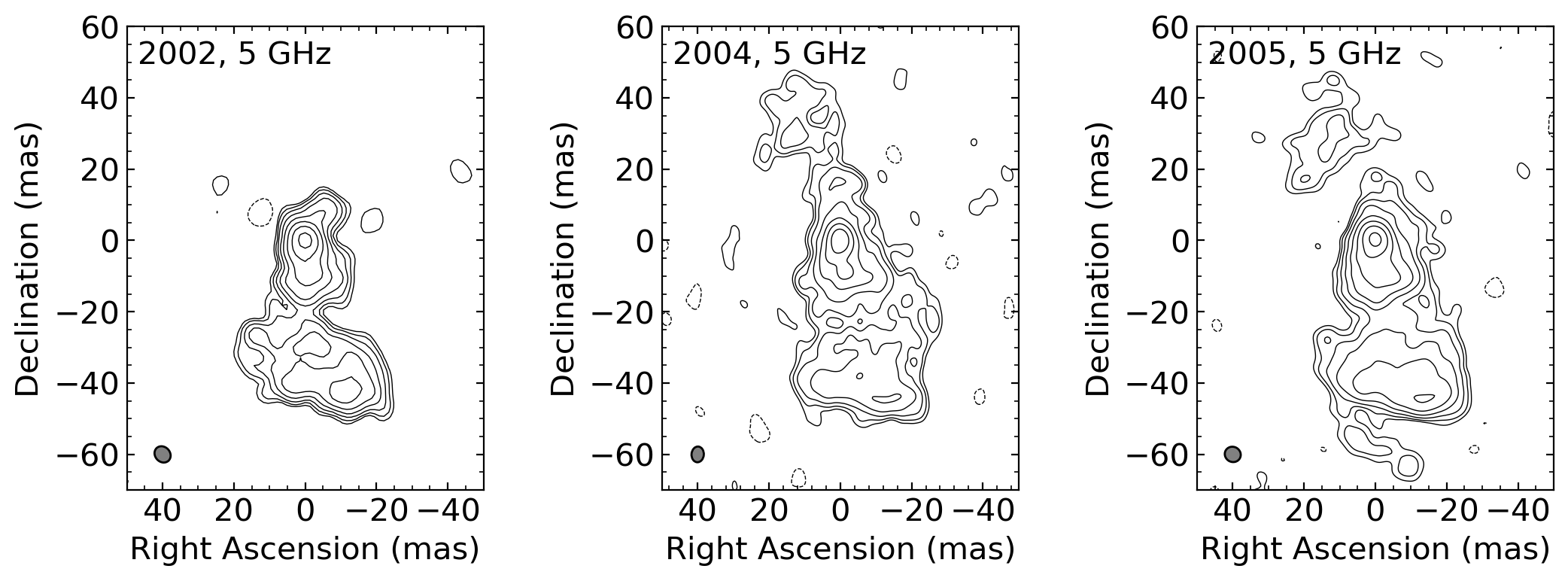}
     \caption{VLBA flux density images of 0954+556 at 5\,GHz. Left panel: Reconstructed image of the visibilities from a combination of data on 20 July 2002 and 22 September 2002. Middle panel: Reconstructed image of the visibilities from a combination of data on 14 June 2004 and 3 November 2004. Right panel: Reconstructed image of the visibilities from a combination of data on 4 April 2005 and 18 August 2005. The contours are drawn at -1, 1, 2, 4, 8, ..., of the first contour level (3$\sigma_\mathrm{rms}$). The synthesized beams are plotted in the bottom left corner of each image.}
    \label{fig:5ghz-com}
 \end{figure*}

\begin{table*}
    \centering
    \caption{Image parameters of concatenated 5 GHz VLBA datasets}
    \label{tab:5ghz-img}
    \begin{threeparttable}
    \begin{tabular*}{0.76\linewidth}{ccccccccc}
    \hline
    Year & Obs. Code  & Date  & $B_{\mathrm{maj}}$ & $B_{\mathrm{min}}$ & $B_{\mathrm{pa}}$ & $\sigma_{\mathrm{rms}}$ & $S_{\text{peak}}$ & $S_{\text{int}}$ \\
     & &   (dd/mm/yyyy) & (mas) & (mas) & ($^\circ$) & (mJy/beam) & (Jy/beam) & (Jy)\\ 
    (1) & (2) & (3) & (4) & (5) & (6)& (7)& (8)& (9)\\
    \hline
    2002 & BH096$^{*}$  & 20/07/2002  & 4.87 & 4.18 & 43.3 & 0.3 & 0.16 & 1.10\\
    & BM170  & 22/09/2002  &  &  &  &  &  & \\
    2004 & BM208A & 14/06/2004  & 4.40 & 3.45 & -4.3 & 0.4 & 0.13 & 1.26 \\
    & BM208B  & 03/11/2004  &  &  &  &  &  & \\
    2005 & BM208C  & 04/04/2005  & 4.54 & 4.24 & 62.41 & 0.3 & 0.15 & 1.27\\
    & BM208D  & 18/08/2005  &  &  &  &  &  &\\
     \hline
    \end{tabular*}
    \begin{tablenotes}[flushleft] 
    \footnotesize  
    \item \textbf{Notes.} The columns are as follows: (1) observation year of concatenated data; (2) observation code of each data point used for concatenation; the data marked with an asterisk represent our observation; (3) observation epoch; (4) to (6) major axis of restoring beam, minor axis of restoring beam, and position angle of major axis, respectively; (7) rms noise level of an image; (8) peak intensity of an image; (9) total flux density of an image.
\end{tablenotes}
\end{threeparttable}
\end{table*}

 \begin{figure*}
 \centering
 	\includegraphics[width=0.8\textwidth]{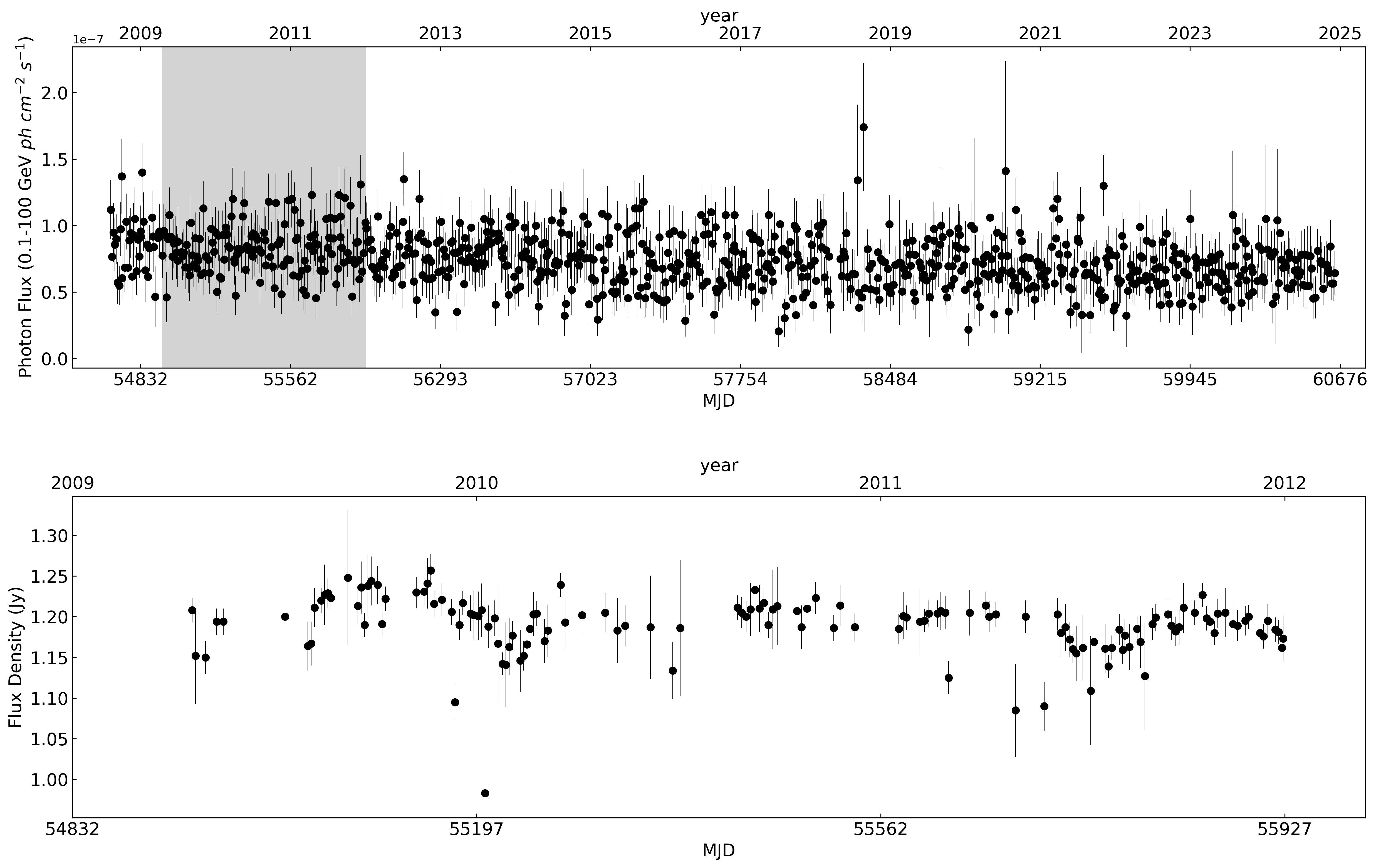}
     \caption{Light curve of 0954+556. The top panel shows a Fermi-LAT $\gamma$ -ray light curve of 0954+556 from 8 August 2008 to 6 December 2024, divided into seven-day bins. All points are plotted along with their statistical errors. The bottom panel is the radio light curve obtained from OVRO at 15\,GHz, from 19 April 2009 to 30 December 2011, with an interval of four days. The gray area in the top panel corresponds to the bottom panel.}
     \label{fig:gamma-light}
 \end{figure*}

\section{Results}
\label{sec:res}
\subsection{Morphology}
\label{subsec:mor}
Figure~\ref{fig:vla_figure} shows the kpc-scale morphology of 0954+556 at 1.4, 5, 8.4, 15, and 22 GHz.
At 1.4\,GHz,  
two extended structures are observed to the northeast and northwest of the brightest emission region (hereafter component NE, component NW, and kpc-scale core, respectively), resembling a triple structure. 
The clear triple structure is revealed by 5 and 8.4\,GHz images with higher angular resolution. 
The northwestern extension is resolved into a linear jet structure (hereafter referred to as linear jet) extending at a position angle of $-60\degr$ for about 3 arcsec,
apparently brightened at its end. The northeastern extension is resolved into an individual knot located about 2 arcsec away from the kpc-scale core. 
On our 15 and 22\,GHz VLA images, the kpc-scale core and the innermost 0.5 arcsec of the jet were detected.
The end of the jet was only detected as diffuse emissions. The rest of the jet and the northeastern knot were resolved and were not detected due to the sensitivity limitation. 

In addition, the images at 1.4 and 5\,GHz are likely to reveal a counter-jet. First, we performed \texttt{MODELFIT} in \texttt{Difmap} to fit 1.4\,GHz data with several circular Gaussian models, which are superimposed on the contours in Figure~\ref{fig:vla_figure}. Table~\ref{tab:modelfit} lists the parameters of each Gaussian, with uncertainties estimated following \cite{2008AJ....136..159L}. Notably, a circular Gaussian located southeast of the core may signal a counter-jet.
Then, at 5\,GHz, a weak emission is detected southeast of the core above 3$\sigma_{rms}$. This feature is considered reliable, as it is clearly visible in the residual map. 
To emphasize the low surface brightness emission, we restored the 5\,GHz image with a circular beam of 900\,mas, approximately twice the size of the initial beam, as presented in the top right panel of Figure~\ref{fig:vla_figure}. With a larger beam, the weak emission exceeds 6$\sigma_{rms}$, enhancing its reliability. 
Therefore, it probably represents the counter-jet at kpc scale.

At pc scales, 0954+556 mainly exhibits two components positioned in the north--south direction, as shown in Figure~\ref{fig:vlba_figure}, consistent with previous studies. 
According to \citet{2005A&A...434..449R}, the two components were named component N and component S, and there was no indication of where the core is. In addition, \citet{2005A&A...434..449R} detected the diffuse tail-like emission (hereafter referred to as tail) extending from component S toward the west at 1.7\,GHz. In our VLBA image at 1.7\,GHz, the tail is traced farther, extending more than 100\,mas to the northwest, roughly aligned with the kpc-scale linear jet. We also detected some emission extending northeastward from component N for a few tens of mas, likely an indication of a counter-jet (CJ). These structures are more clearly revealed by the slightly tapered image (top right panel of Figure\,\ref{fig:vlba_figure}).

In the VLBA images at 5\,GHz, \cite{2005A&A...434..449R} and \cite{2011ApJ...738..148M} detected component N and component S, while our observation revealed a greater number of faint features.
As mentioned in Section~\ref{ssec:obs}, we accumulated seven epochs of 5\,GHz VLBA data over a period of 14 years from July 2002 to July  2016.
In Figure\,\ref{fig:vlba_figure}, we present the image of the latest epoch with the highest dynamic range. Due to an on-source time of $\sim$3 hours, the CJ was more clearly detected, extending first to the northwest and then bending to the northeast. The bright portion of the tail was also detected. Because of the limited on-source time and the lack of antennas from other 5\,GHz datasets, to reconstruct images with higher sensitivity we concatenated $uv$ data from 20 July 2002 and 22 September 2002; from 14 June 2004 and 3 November 2004; and from 4 April 2005 and 18 August 2005. Since the source is stable over seven epochs, and the two $uv$ datasets used for concatenation in each year are close in time, our results are reliable. Three reconstructed images (in 2002, 2004, and 2005) are shown in Figure\,\ref{fig:5ghz-com}, and the image parameters are listed in Table\,\ref{tab:5ghz-img}. In Figure\,\ref{fig:5ghz-com}, the image in 2002 only show the emission of the CJ extends northwest, while the images in 2004 and 2005 show the CJ extending northwest and bending northeast, consistent with the image in 2016. Thus, we consider the reliable CJ to be the counter-jet at pc scale.

At 8.4~GHz,  
the brightest emitting region in component N elongates in the north--south direction and possibly turns southwestward, indicating an inner jet in this region. 
Most of the emissions from component S were resolved out, and only the brightened region and its southern edge were detected, implying a spectral steepening between 5 and 8.4\,GHz in this component. The Radio Fundamental Catalog (RFC) provides the highly accurate coordinates of distant extragalactic radio sources derived from X--S dual-frequency astrometric analysis, helping build the International Celestial Reference Frame (ICRF). As 0954+556 is registered with ICRF, the image center of 0954+556 is aligned with the position provided by RFC. We used the latest position from RFC (rfc$\_$2025a,\footnote{\url{https://astrogeo.smce.nasa.gov/sol/rfc/rfc_2025a/}} \citealt{2025ApJS..276...38P}) to roughly
represent the brightest position in the 8.4\,GHz VLBA image, i.e., the image center, and used the optical position in Gaia Data Release 3\footnote{\url{https://gea.esac.esa.int/archive/}} \citep{2023A&A...674A...1G}. Following \cite{Gaia-VLBI}, the angular offset between the Gaia and VLBI positions is 2.58$\pm$0.22 mas. This small offset supports the identification of component N as the region with the radio core, and the CJ component as the counter-jet.

The innermost core region of 0954+556 is revealed at higher frequencies. The 15\,GHz VLBA image in \citep{2005A&A...434..449R} displayed a well-defined jet in component N, initially extending southeast before bending southwest. In our work, 0954+556 is imaged for the first time at 22 and 43\,GHz at pc scale. The 22\,GHz map produced with data in the $uv$ range between 2.3 and 48.0\,$\mathrm{M}\lambda$ is presented in the lower left panel of Figure~\ref{fig:vlba_figure}. The structure is consistent with the bending observed at 15\,GHz.
The lower right panel of Figure~\ref{fig:vlba_figure} shows the tentative 43\,GHz image of 0954+556.
The detected emission is compact, slightly extending southward, and can be fitted with a circular Gaussian with a radius $\sim78\mu\mathrm{as}$ (0.624\,pc). 

The jet morphology of 0954+556 exhibits a complex bending pattern both at pc and kpc scales. 
At pc scales, the jet initially extends southeast and then bends southwest, as shown in the 15 and 22\,GHz VLBA images. The lower frequency images (1.7, 5, and 8.4\,GHz) indicate that the jet possibly bends back toward the southeast, reaching the edge of component N, and continues southwestward, connecting to component S. The pc-scale counter-jet initially extends northwest, then turns toward the northeast. To connect the structure from pc to kpc scales, the observed tail emission might suggest that the pc-scale forward jet undergoes a single bend toward the northwest to align with the kpc-scale forward jet. However, it seems that the pc-scale counter-jet possibly bends southeast to align with the kpc-scale counter-jet, and then bends again to the north to reach component NE.

\subsection{Variability}
\label{subsec:var}

Figure \ref{fig:gamma-light} shows the $\gamma$-ray light curve
and the 15\,GHz radio light curve of 0954+556 produced with the data introduced in Section~\ref{ssec:obs}. No flare is seen in either of the light curves. 
To quantify the variability of 0954+556, we utilized the fractional variability amplitude \citep[$F_{var}$;][]{2002ApJ...568..610E,2003MNRAS.345.1271V}, and the related equations are given here.

The mean square error is defined as
\begin{equation}
\bar{\sigma}_{\mathrm{err}}^2=\frac{1}{N} \sum_{i=1}^N \sigma_{\mathrm{err}, i}^2
\end{equation}
where $N$ is the number of data points, and $\sigma_{\mathrm{err}, i}$ is the error on the data point. In addition, the sample variance is 
\begin{equation}
S^2=\frac{1}{N-1} \sum_{i=1}^N\left(x_i-\bar{x}\right)^2
\end{equation}
where $x_i$ is the value of the data point, and $\bar{x}$ is the mean of $x_i$. Therefore, the fractional variability amplitude is given by
\begin{equation}
F_{\mathrm{var}}=\sqrt{\frac{S^2-\bar{\sigma}_{\mathrm{err}}^2}{\bar{x}^2}}
\end{equation}
and the uncertainty of fractional variability amplitude $\Delta F_{\mathrm{var}}$ is calculated as
\begin{equation}
\operatorname{err}\left(\sigma_{\mathrm{NXS}}^{2}\right)=\sqrt{\left(\sqrt{\frac{2}{N}} \frac{\bar{\sigma}_{\mathrm{err}}^2}{\bar{x}^2}\right)^2+\left(\sqrt{\frac{\bar{\sigma}_{\mathrm{err}}^2}{N}} \frac{2F_{\mathrm{var}}}{\bar{x}}\right)^2}
\end{equation}
\begin{equation}
    \Delta F_{\mathrm{var}}=\sqrt{F_{\mathrm{var}}^{2}+\operatorname{err}\left(\sigma_{\mathrm{NXS}}^{2}\right)}-F_{\mathrm{var}}
\end{equation}

The $F_{var}$ derived from the LCR dataset and the 15\,GHz OVRO data is 6.9$\pm$2.5\% and  1.6$\pm$0.3\%, respectively, and are considerably lower than those observed in blazars, which are known for their strong variability. For example, in the LCR dataset, the blazar PKS\,0805-07 exhibits an $F_{var}$ above 75\% \citep{2024ApJ...977..111A}, and the blazar 3C\,279 shows an $F_{var}$ of 100\% \citep{2018ApJ...858...80R}. In the radio band, the $F_{var}$ of blazar Mrk421 obtained by OVRO is 15\% \citep{2019Galax...7...62S}, while blazar 3C\,273 reaches an $F_{var}$ of 29\% at 15\,GHz from the data from the University of Michigan Radio Astronomy
Observatory (UMRAO) \citep{2008A&A...486..411S}. 
Compared with typical blazars, 0954+556 is not significantly variable, thus indicating stable emission zones for 0954+556. Notably, although 0954+556 is classified as a FSRQ in the Fermi catalog, \cite{2020arXiv200511208B} reported that while the blazars all have fractional variability exceeding 10$\%$, 0954+556 shows a flux decrease of about 13$\%$ from August 2008 to August 2018, without any flare.
To assess the stability of the structure, we analyzed all the total flux density images at 5\,GHz, from 2002, 2004, 2005, and 2016. All of the 5\,GHz images were convolved with a common beam corresponding to the observation in 2016, and 
an average image was constructed by taking the mean of the four images. The residual maps were produced by subtracting each individual image from the average, as shown in Figure~\ref{fig:subtract}. The intensity differences across these residuals range from -15 mJy/b to 15 mJy/b, which is negligible compared to the intensity of the source. Thus, the structure of 0954+556 remained stable over the 14-year span, without significant variability.

  \begin{figure*}[h!]
  \centering
 	\includegraphics[width=0.7\textwidth]{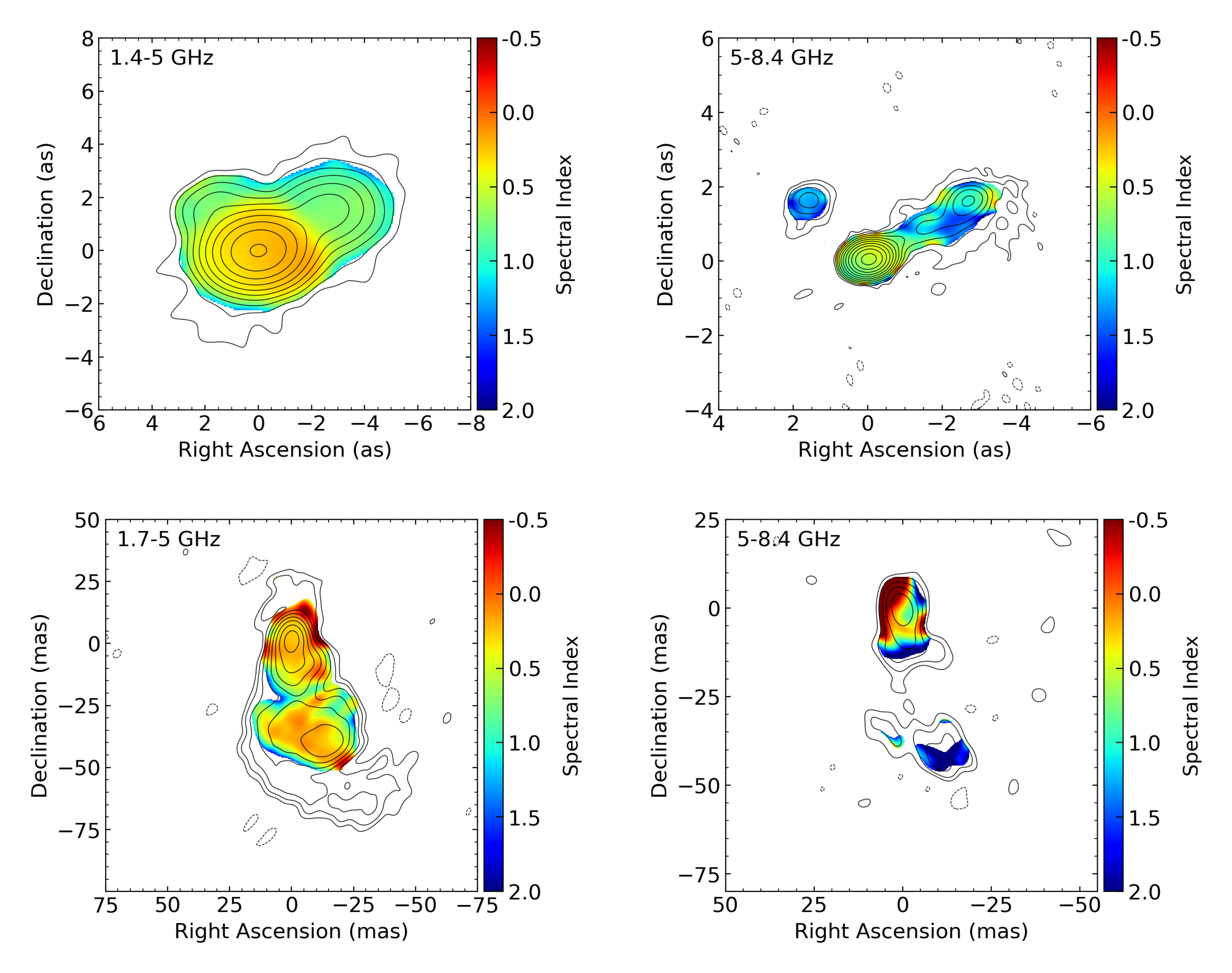}
     \caption{Spectral index distribution maps of 0954+556. From left to right, the panels show the distribution of spectral index between 1.4 and 5\,GHz at kpc scales, between 5 and 8.4\,GHz at kpc scales, between 1.7 and 5\,GHz at pc scales, and between 5 and 8.4\,GHz at pc scales. 
     The overlapped contour images are a VLA image at 1.4\,GHz with a $uv$ range $5.1-168.7 \mathrm{K}\lambda$,
     a VLA image at 5\,GHz with a $uv$ range $11.3-561.2 \mathrm{K}\lambda$, a VLBA image at 1.7\,GHz with a $uv$ range $2.3-48.0 \mathrm{M}\lambda$, and a VLBA image at 5\,GHz with a $uv$ range $5.4-39.5 \mathrm{M}\lambda$, respectively. All counters start at 3$\sigma_\mathrm{rms}$ and increase by a factor of 2.
     The color represents the value of spectral index $\alpha$. 
     Pixels with flux density below 5.4$\sigma_\mathrm{rms}$, alongside those with spectral index uncertainty exceeding 40\%, are blanked.}
     \label{fig:index}
 \end{figure*}

 \begin{figure*}[h!]
 \centering
 	\includegraphics[width=1.0\textwidth]{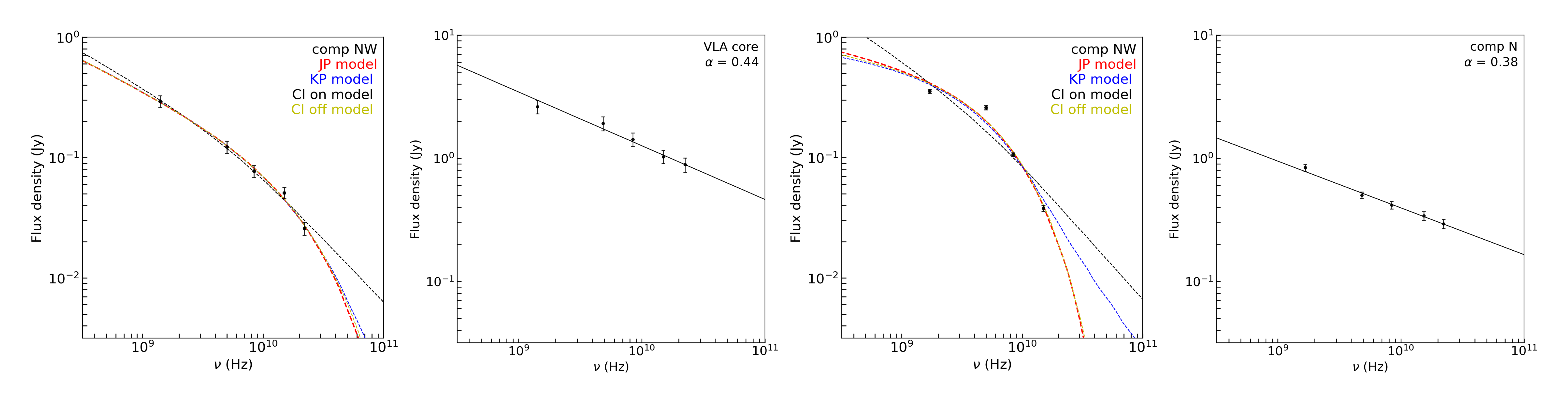}
     \caption{Four different fit results for each component of 0954+556. The red, blue, black, and yellow dashed lines represent the JP, KP, CI-on, and CI-off models respectively. The VLA core and component N are fitted by a simple power law. For component N, the data at 43 GHz was not used in the spectral analysis as mentioned in Section~\ref{sssec:prsp}.}
     \label{fig:rad-age}
 \end{figure*}
 
\section{Spectral analysis}
\label{sec:spa}
\subsection{Spectral index distribution}
\label{ssec:sp_kpc_pc}

The kpc-scale spectral index maps were derived from the VLA images at 1.4 and 5 GHz and 5 and 8.4\,GHz, respectively. 
At 1.4 and 5\,GHz, the NE component was fitted with circular Gaussian models with flux densities of 61.4$\pm$14 mJy and 24.6$\pm$5 mJy, respectively, yielding $\alpha$$\sim$0.7$\pm$0.2 and suggesting an optically thin synchrotron emission between the two frequencies.
Analogously, at 5 and 8.4\,GHz, the NE component was modeled with circular Gaussian models of $22.5\pm4.3$ and $10.5\pm3.1\mathrm{mJy}$, respectively, resulting in $\alpha$$\sim$1.5$\pm$0.7, further supporting its optically thin nature. 
Thus, its positions measured at each frequency pair should be identical with each other \citep{2008A&A...483..759K}.
We therefore aligned the images by the centroid of component NE to produce kpc-scale spectral index maps. 

The top panel of Figure~\ref{fig:index} shows the spectral index distribution at kpc scales,
overlaid with contours representing the 1.4 and 5\,GHz VLA images.
In the 1.4--5\,GHz spectral index map, the spectral index distribution of the core, the component NE and NW tend to be flat, with $\alpha$ above 0. While the component NE and NW tend to have approximate spectral index distribution, the core displays a flatter spectral index.  
In the 5--8.4\,GHz spectral index map, the steepest spectrum appears in component NE.
The spectrum at the kpc-scale core remains flat.
Going outward along the linear jet, the spectrum steepens and the $\alpha$ varies. At the jet terminus, the spectrum is flattened, and the $\alpha$ decreases.
The brightening and spectral flattening in the jet terminus indicate a local acceleration of emitting particles. 

At pc scales, we used 1.7 (7 February 2000), 5 (20 July 2002), and 8.4 (20 July 2002) \,GHz VLBA images to produce spectral index maps.
Since no compact optically thin component was identified at the pc scales, we utilized a 2D cross-correlation method \citep{2013A&A...557A.105F,2014JKAS...47..195K} to determine the core-shift between two frequencies, and thus to align the images. Given that the 1.7 GHz and 5 GHz observations are not simultaneous, we attempted to construct a spectral index map by combining the 1.7\,GHz data (7 February 2000) with three 5\,GHz datasets (20 July 2002, 3 November 2004, and 4 April 2005). The three resulting spectral index maps are very similar, and thus we chose the closest epoch. In addition, the source has very weak variability according to Section~\ref{subsec:var}, making our spectral index map reliable and convincing.
The bottom panels of Figure~\ref{fig:index} present the 1.7--5\,GHz spectral index map 
overlapping on the 1.7\,GHz VLBA contour image and the 5--8.4\,GHz spectral index map 
overlapping on the 5\,GHz VLBA contour image.

The 1.7--5 GHz spectral index map shows that the spectral index distribution of components N and S are very similar; both exhibit flat spectra across most of their regions and steeper spectra toward the edges. This similarity makes it challenging to identify the core region based solely on the spectral index map at these frequencies. However, in the 5--8.4 GHz spectral index map, component S shows a significantly steeper spectrum with $\alpha \sim2$, while component N retains a relatively flat region, suggesting continued particle acceleration in component N rather than S.

\subsection{Radiative age}
\label{ssec:ra}

An aging synchrotron spectrum from a relativistic electron population with an initial injected power-law energy distribution is described as a power-law  
steepening at high frequencies due to particle energy losses \citep{1962SvA.....6..317K,1970ranp.book.....P,1999A&A...345..769M}.
The break frequency where the spectrum starts steepening depends on the elapsed 
time since the particles were injected, and on the intensity of the magnetic field.

To perform spectral aging analysis, we utilized a Python-based package \texttt{SYNCHROFIT}\footnote{\url{https://github.com/synchrofit/synchrofit}}\citep{2018MNRAS.474.3361T,2018MNRAS.476.2522T,2022MNRAS.514.3466Q}, 
which uses an adaptive maximum likelihood algorithm to fit the observed radio spectrum. 
\texttt{SYNCHROFIT} provides continuous-injection on (CI-on), continuous-injection off (CI-off), Jaffe-Perola (JP), and Kardashev-Pacholczyk (KP) models \citep{1962SvA.....6..317K,1970ranp.book.....P,1973A&A....26..423J,1999A&A...345..769M,2007A&A...461..923O} to estimate the radiative age of jet components. 
In the CI-on and CI-off models the electron population is injected continuously, so electrons with different ages are mixed. 
The injection keeps taking place in the CI-on model, 
whereas the continuous injection was kept for a long time and switched off in the CI-off model.
In the JP and KP models the particles were injected in a single pass. The difference between the JP and KP models is whether electron pitch-angle scattering is considered or not. 
Among the four models, the break frequency $\nu_\mathrm{br}$ and the non-aged spectral index $\alpha_\mathrm{inj}$ are constrained by fitting the flux density across multiple frequencies.
In addition, the CI-off model gives the remnant fraction T, i.e., the fractional time spent in an inactive phase to the total source age.

The radiative age $\tau_\mathrm{syn}$ is derived as
\begin{equation}\label{eq:tau_syn}
\tau_{\text {syn}}=\frac{v B^{1 / 2}}{B^2+B_{\text {IC}}^2}\left(\nu_{\text {br}}(1+z)\right)^{-1 / 2},
\end{equation}
in which $z$ is the redshift of the source, $B$ is the magnetic field, $B_{\mathrm{IC}}=3.18\times10^{-6}(1+z)^2$ gauss is the magnetic field equivalent to the microwave background, and $v$ is a proportional constant. The details are in Section 2.2.1 of \citet{2018MNRAS.474.3361T}.

We used the equipartition magnetic field strength ($B_\mathrm{eq}$) representing the condition of minimum total energy density as the magnetic field to derive $\tau_\mathrm{syn}$. The following equation given by \cite{1980ARA&A..18..165M} was adopted to calculate the magnetic field strength,
\begin{equation} \label{eq:beq}
\begin{aligned}
B_{\text{eq}}=5.69 \times 10^{-5}\left[\frac{(1+\text{k})}{\eta}(1+\text{z})^{3-\alpha}\right.  \frac{1}{\theta_x \theta_y \text{~s} \sin ^{3 / 2} \phi} \\
 \left.\times \frac{F_0}{v_0^\alpha} \frac{v_2^{\alpha+1 / 2}-v_1^{\alpha+1 / 2}}{\alpha+\frac{1}{2}}\right]^{2 / 7} \text {gauss}.
\end{aligned}
\end{equation}
Here $\text{k}=1$ is the ratio of heavy particle energies to electron energies, $\eta=1$ is the filling factor, and $z$ is the redshift of a target. 
Based on assuming cylindrical symmetry, $\theta_x$ and $\theta_y$ represent the size of a component in a plane perpendicular to the line of sight, and $s$ is the length of a component along the line of sight; $\phi$ is the angle between the magnetic field and the direction of the line of sight; $F_0$ in Jy is the flux density at frequency $v_0$; and $\alpha$ is the spectral index. 

The model assessment is conducted by a reduced-chi square equation,
\begin{equation}
\chi_{\text{red}}^2=\frac{1}{(N-P)} \sum_{i=1}^N\left(\frac{x_i-\mu_i}{\sigma_i}\right)^2,
\end{equation}
where $N$ is the number of observation data points, $P$ is the number of fitted parameters, $N-P$ represents the degree of freedom, $x_i$ are the observed values,  
$\mu_i$ are the expected values based on the model, and $\sigma_i$ is the uncertainty of the observed values. 

The integral flux density of each component was extracted with \texttt{AIPS} task \texttt{TVSTAT}.
Its uncertainty was estimated as $\mathrm{rms}*N_{\mathrm{region}}/\sqrt{N_{\mathrm{beam}}}$ \citep{2021A&A...654A..27B}, where $N_{\mathrm{region}}$ is the number of pixels of the chosen emission region, and $N_{\mathrm{beam}}$ is the number of pixels of the resolution beam.
The estimated flux density with the uncertainty of each component is listed in Table~\ref{tab:flux-of-spectral-analysis}.

\subsubsection{Spectral aging at  kpc scale}
\label{sssec:ra_kpc}
Section~\ref{sssec:prsp} described how we prepared the images for spectral fitting, and the resultant images are displayed in Figure~\ref{fig:rad-age-vla}. After being convolved with the beam at the lowest frequency, the linear jet becomes a circular Gaussian, so the overall structure can be recognized as the core, component NE, and component NW (Figure~\ref{fig:rad-age-vla}).
The fitted values of $\alpha_\mathrm{inj}$, $\nu_\mathrm{br}$, 
and T are listed in Table\,\ref{tab:rad-age}.
The fitting curves overlapped by data points are presented in Figure\,\ref{fig:rad-age}.

The flux density of the core component is well fit by a simple power law with an index of $\sim$0.44. For component NW, the fitted injection index from the CI-on model is $\alpha_\mathrm{inj}=$0.54, while the other models yield a slightly lower value of $\alpha_\mathrm{inj}=$ 0.49. The CI-on and the KP models both give a $\nu_\mathrm{br}$ within the observed frequency range, whereas the CI-off and JP models produce a $\nu_\mathrm{br}$ that is a bit higher than the highest-frequency point. The quality of fits is good, with the $\chi_{\text{red}}^2$ of 1.1 for the CI-on model and $\sim$0.6 for the other models.
To estimate $B_{\text{eq}}$, according to the morphology in Figure~\ref{fig:rad-age-vla}, we assumed a circular geometry for component NW with a diameter of $3.5 \, \mathrm{arcsec}$, resulting in the corresponding magnetic field strength as $4.4\times10^{-5}$ gauss.
The resultant $\tau_\mathrm{syn}$ for component NW ranges from 0.3 to 1.4 \,$\mathrm{M}$yr.
It is notable that above 15\,GHz, the emissions of NW are dominated by the hot spot at the jet terminus, and thus the derived radiative age should be much lower than the actual source age. 
Due to the lack of data points, fitting the spectrum of component NE was not performed.

\subsubsection{Spectral aging at pc scale}
\label{sssec:ra_pc}
The integral flux densities with uncertainties of component N and component S were extracted in the same way, as described in Section~\ref{sssec:prsp} and Section~\ref{ssec:ra}. The derived values and images are listed in Table~\ref{tab:flux-of-spectral-analysis} and Figure~\ref{fig:rad-age-vlba}.
The fitted values of $\alpha_{\mathrm{inj}}$ and $\nu_\mathrm{br}$ are listed in Table\,\ref{tab:rad-age}.
The fitting curves overlapped by data points are presented in Figure\,\ref{fig:rad-age}.

The emission of component N between 1.7 and 22\,GHz can only be fit with a simple power law with an index of 0.38. 
The observational spectrum of component S is curved, and apparently our data disfavor the CI-on model as the CI-on model steepens too gradually. Among all models, the CI-on model yielded the highest $\chi_{\text{red}}^2$, more than twice that of the other models.
For the remaining models, the fitting values of $\nu_\mathrm{br}\sim5\mathrm{GHz}$ are reasonable since they are consistent with what is revealed by the total intensity maps. 
Coupled with the brightening at the southwestern edge of component S below 8.4\,GHz, a local acceleration in this region is strongly indicated.
Furthermore, the spectrum steepening sharply beyond 5\,GHz suggests that the acceleration might be historical and had been quenched for a long time.

The geometric size of component S was estimated using a circular region with a diameter of 40\,mas (Figure~\ref{fig:rad-age-vlba}). The magnetic field strength in component S was estimated to be $2.6\times10^{-3}$ gauss, assuming energy equipartition.
Therefore, in the CI-off, JP and KP models the value of $\tau_\mathrm{syn}$ was derived to be $\sim3.6\,\mathrm{kyr}$, $\sim2.7\,\mathrm{kyr}$, and $\sim1.8\,\mathrm{kyr}$, respectively.
The actual age of S should be older than these estimated radiative ages, considering the historical local acceleration in this region. The separation between the brightest part of component N and component S is $\sim$45\,mas, i.e., 360\,pc. Assuming a typical advance speed of 0.03-0.1$c$ for CSOs, the estimated dynamic age of component S is 11.7--39.1 $\mathrm{kyr}$, close to its radiative age.

\begin{table*}
    \centering
    \caption{Total flux density of each component for spectral analysis}
    \label{tab:flux-of-spectral-analysis}
    \begin{tabular*}{0.68\linewidth}{cccccc}
    \hline
    Comp  & 1.4 GHz & 5 GHz& 8.4 GHz & 15 GHz & 22 GHz \\
       & (mJy) &(mJy) &(mJy) &(mJy) &(mJy)\\
    \hline
     VLA core  & 2622.2$\pm$326 & 1918.3$\pm$249 & 1423.6$\pm$183 & 1028.3$\pm$127 & 887.8$\pm$117  \\
    NE   & 65.8$\pm$7.8   &25.3$\pm$3.4    &13.7$\pm$2.2  &              & \\
    NW & 293.7$\pm$32.4 & 122.9$\pm$14.4 & 77.1$\pm$8.7 & 51.0$\pm$5.5 & 25.8$\pm$3.2 \\
    \hline 
    Comp & 1.7 GHz & 5 GHz& 8.4 GHz & 15 GHz & 22 GHz \\
      &(mJy) & (mJy) &(mJy) &(mJy) &(mJy) \\
    \hline
    N & 838.3$\pm$51.3 & 501.1$\pm$30.8 & 415.9$\pm$29.5 &341.1$\pm$27.3 & 293.2$\pm$25.4 \\
    S & 354.6$\pm$14.4  & 260.8$\pm$11.0& 106.0$\pm$3.6 & 37.9$\pm$2.2&  \\
    \hline
\end{tabular*}
\end{table*}

\begin{table*}
    \centering
    \caption{Results of radiative age}
    \label{tab:rad-age}
    \begin{threeparttable}
    \begin{tabular*}{0.65\linewidth}{cccccccc}
    \hline
    Comp &$B_{\mathrm{eq}}$ & model & $\alpha_{\text {inj }}$ & $\nu_{\mathrm{br}}$ &$ \tau_{\mathrm{syn}}$  & T &$\chi_{red}^2$ \\
      &($10^{-5}$ gauss) & &  & (GHz)    &($\mathrm{kyr}$) & &\\
     (1)&(2)&(3)&(4)&(5)&(6)&(7)&(8)\\ 
    \hline
    VLA core & & power law & 0.44& &  & &0.45\\
    NW& 4.4 & CI-on& 0.54& 6.9 & 1401 & & 1.1\\
      &     & CI-off& 0.49 & 26.4 & 719 & 0.85  & 0.58 \\
    & &  JP& 0.49 & 30.4 &  671 & & 0.61\\
    & & KP& 0.49 & 18.2 & 385 & & 0.64\\
    \hline 
    N & & power law & 0.38 & & & &0.45 \\
    S & 263.7  & CI-on& 0.62 & 5.8 & 3.5 & & 75.6 \\
      &     & CI-off& 0.25 &  5.7 & 3.6 &0.78  & 24.1\\
    & & JP& 0.25 & 9.8 & 2.7 & & 25.3\\
    & & KP& 0.19 & 4.6 &1.8 & & 29.7\\
     \hline
    \end{tabular*}
    \begin{tablenotes}[flushleft] 
    \footnotesize  
    \item \textbf{Notes.} The columns are as follows: (1) components of the target source; (2) equipartition magnetic field strength of each component; (3) model used to fit the observed radio spectra; (4) spectral injection index; (5) break frequency obtained from spectral analysis; (6) radiative age of each component; (7) remnant fraction, i.e., the fractional time spent in an inactive phase to the total source age; (8) reduced chi-squared values refer to different models.
\end{tablenotes}
\end{threeparttable}
\end{table*}

\begin{table}
    \centering
    \caption{Parameters of polarization images}
    \label{tab:polarization-para}
    \begin{threeparttable}
    \begin{tabular*}{0.73\linewidth}{cccccc} 
    \hline 
    Freq & $P_{\mathrm{tot}}$ &$P_{\mathrm{peak}}$ &$\sigma_{Q, U}$ & $m$\\
    (GHz)               &(mJy)              & (mJy/b)        &(mJy/b)              & (\%)\\
    (1) & (2) & (3) & (4) & (5)   \\ 
    \hline
    1.7 & 77.0 &14.4  & 0.4 & 3.8\\
    5  &51.3   & 11.1 & 0.3 & 5.2\\
    8.4 &40.0  & 8.7  & 0.3 & 7.5\\
    \hline 
    \end{tabular*}
    \begin{tablenotes}[flushleft] 
    \footnotesize  
    \item \textbf{Notes.} The columns are as follows: (1) observational frequency; (2) total flux density of cleaned $P$ images, $P_{\mathrm{tot}}=\sqrt{U^{2}+{Q}^2}$; (3) peak flux intensity of the polarization image; (4) rms noise level of $P$ image ($ \sigma_{P}=(\sigma_{U}+\sigma_{Q})/2$); (5) average fractional polarization ($\overline{m}=P_{\mathrm{tot}} / I_{\mathrm{tot}}$)
    \end{tablenotes}
    \end{threeparttable}
\end{table}

 \begin{figure*}[h!]
 \centering
 	\includegraphics[width=1\textwidth]{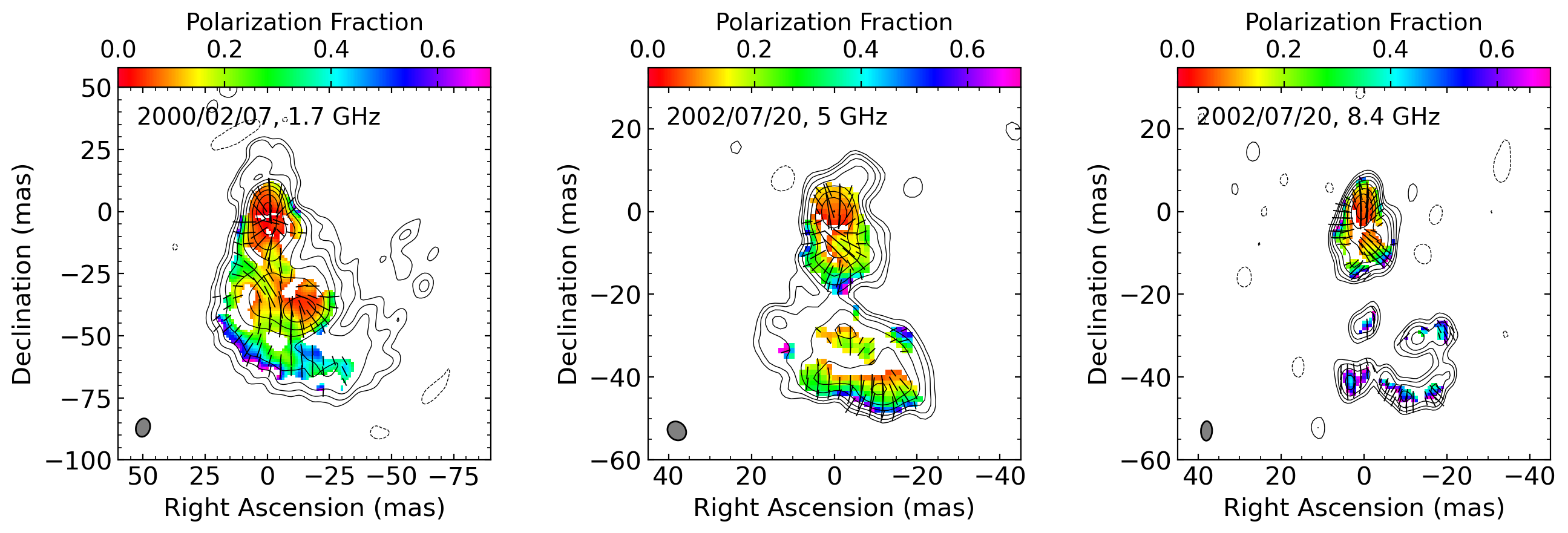}
     \caption{Polarization images at different frequencies. The restoring beam is indicated by a gray ellipse in the lower left corner of each image. The contours are drawn at $-$1, 1, 2, 4, 8, ..., of the first contour level (3$\sigma_\mathrm{rms}$). Superimposed on the contours of flux density are vectors whose lengths are proportional to the polarized intensity; the directions indicate the direction of the $\textit{E}-$vectors, and the colors show the fractional linear polarization. In all polarization images, the 1 mas line is 1 mJy/b. We clipped pixels with $I<3\sigma_{I}$, $P<3\sigma_{P}$, and $m<3\sigma_{m}$.}
     \label{fig:pola}
 \end{figure*}

\section{Polarization}
\label{sec:pol}
The 1.7\,GHz VLBA observation conducted in February 2000,
and the 5 and 8.4\,GHz VLBA observations conducted in July 2002 were designed for polarimetry (Section~\ref{ssec:obs}). 
The results of the polarimetry are presented in Figure~\ref{fig:pola}. The fractional polarization ($m=P/I$) is shown as pseudo-color, overlapping contours of the total intensity.
The observed EVPAs are indicated by vectors. Following the description in \cite{2012AJ....144..105H}, we estimated the uncertainties of Stokes I, Q, U images, and considered the errors associated with antenna feed leakage. Then we estimated the uncertainties of $P$, $m$, and EVPA, and clipped pixels with $I<3\sigma_{I}$, $P<3\sigma_{P}$, and $m<3\sigma_{m}$, as described in \cite{2024ApJ...974..111H}. Table~\ref{tab:polarization-para} lists the statistics of the above images: total polarized flux density, peak polarized intensity, rms noise of the polarization image, and the average fractional polarization ($\overline{m}=P_{\mathrm{tot}} / I_{\mathrm{tot}}$). 
From Figure~\ref{fig:pola} we found that the EVPA distributions at 5 and 8.4\,GHz are nearly identical, and its RM at kpc scales was estimated to be $+1\pm1\,\mathrm{rad}\,\mathrm{m}^{-2}$ \citep{1981ApJS...45...97S}, allowing us to obtain the EVPA distribution at 1.7\,GHz, as mentioned in Section~\ref{ssec:polc}.

0954+556 is a rare case in that the detected polarized emissions cover most of the jet structure at pc scales. In general, the jet at pc scales propagates toward the south (from component N to S) and then bends to the northwest (the tail).
The polarization structure of 0954+556 reveals a clear spine-sheath configuration throughout the entire structure \citep{2005MNRAS.356..859P,2014MNRAS.444..172G,2021Galax...9...58G}. In the central spine region, the EVPAs are aligned with the jet axis, whereas the EVPAs are roughly perpendicular to the jet direction at the edges, i.e., the sheath. This means the inferred B-field at the inner jet is oriented orthogonally to the jet and at the jet edges along the jet \citep{1970ranp.book.....P}, which is a strong indication of the existence of the helical magnetic field at pc scales within the jet \citep{2021Galax...9...58G}.
Notably, the EVPAs in the spine region basically follow the jet even at the bending, further supporting the idea that the jet has bent a few times to reach the brightest region of component S, as mentioned in Section~\ref{subsec:mor}.

The average fractional polarization $\overline{m}$ is 3.8$\%$, 5.2$\%$, and 7.5$\%$ at 1.7\,GHz, 5\,GHz, and 8.4\,GHz, respectively. Interestingly, the fractional polarization $m$ increases from the central spine toward the outer sheath. At all three frequencies, component N shows an increase in $m$ from $\sim$0.05 in the central region to $\sim$0.3 at the jet edge, with the localized edge reaching $m\sim0.7$. Component S exhibits the same trend at 1.7 and 5\,GHz as component N, and shows a high $m$ of 0.4 -- 0.7 at 8.4\,GHz.
The tail observed at 1.7\,GHz shows high values of $m$ ranging from 0.3 to 0.7. This is also an indication of the existence of a helical magnetic field \citep{2021Galax...9...58G}.

\section{Discussion and conclusion}
\label{sec:dis}
\subsection{Classification}
As mentioned in Section~\ref{sec:intro}, the classification of 0954+556 is controversial; it is either a FSRQ in the $\gamma$-ray frequency range or a CSO in the radio frequency range. A blazar is usually referred to as an AGN with a powerful relativistic jet aligning well with our line of sight, characterized by extremely high brightness temperatures ($T_\mathrm{b}$) over the threshold of inverse Compton catastrophe ($10^{12}\mathrm{K}$), apparent superluminal motions observed at pc scales, and fast and drastic variations of flux density over a wide range of frequencies \citep{2015ApJ...798..134H,2018ApJ...866..137L,2021ApJ...923...67H}.
However, for 0954+556, the $T_b$ of the brightest emission component detected by VLBA at 5 and 15\,GHz was only at the magnitude of $10^8 \mathrm{K}$ \citep{2005A&A...434..449R, 2011ApJ...738..148M}. Following the method in \cite{2021ApJ...923...67H}, we estimated the $T_b$ of the brightest component at 8.4\,GHz to be $\sim6\times10^8\mathrm{K}$. No superluminal motion was reported. No evidence of variability was found in $\gamma$-rays over the 19-month LAT observing period from August 2008 to March 2010, and its optical flux density was reported to change by only 30\% over 7 years from 2002 to 2009 \citep{2011ApJ...738..148M}, nor was any variability at radio wavelengths found \citep{2007A&A...462..547F}. Our analysis also supports that the variation of 0954+556 is much lower than typical blazars at $\gamma$-ray and radio bands (see Section~\ref{subsec:var}), indicating that 0954+556 is a non-aligned AGN. 

The structure of 0954+556 detected by VLBA resembles a compact double with a projected separation of 360\,pc, leading to 0954+556 constituting a CSO in previous studies. 
However, the core position of this CSO was not identified. In our results, unlike that of component N, the spectral index distribution of component S shows a significant decline in its flux density over 5\,GHz, indicating a difference between the two components. As analyzed in Section~\ref{sssec:ra_pc}, the radiative age analysis reveals that there is no particle acceleration in component S, whereas component N harbors the radio core in which the acceleration of the particles is still taking place. 
There is evidence that the radio core is located in the brightest emission region of component N.
First, the optical position of this source lies just 2.58$\pm$0.22\,mas from the peak of the 8.4\,GHz VLBA image (Section~\ref{subsec:mor}). Given that the angular size of component N is approximately 25\,mas, this close alignment strongly suggests that the optical and radio cores coincide. Second, our observations detected the counter-jet at 1.7 and 5\,GHz (Figure~\ref{fig:vlba_figure}).
Third, a jet extending southward is observed in the central part of the structure (Section~\ref{sec:res} and Section~\ref{sec:pol}). At 15\,GHz, this structure appears as a well-defined jet initially extending southeast and subsequently turning southwest. 
The southeast segment of this jet is also visible at 22\,GHz. The images at 1.7, 5, and 8.4\,GHz show that the jet continues to elongate to the southeast. To connect to the brightest region of component S, the jet then extends southwest again. 
The EVPAs within the jet are along the jet, while they are perpendicular to the direction of jet propagation at the edges of the structure (Figure~\ref{fig:pola}), indicating a spine-sheath polarization structure. Additionally, the fractional polarization increases from the central spine toward the outer sheath, from $\sim$ 0.05 to $\sim$ 0.7. 
Finally, the radio spectrum of component N between 1.7 and 22\, GHz can only be fit by a power law with an index of 0.38 (Section~\ref{sssec:ra_pc}),
implying a continuous supply of relativistic electrons.
Thus, at pc scales we identify that the core of this CSO is located at component N.

Although the radio core is near the position of peak brightness, it is not identified in our VLBA images. The core usually exhibits a flat spectrum, and our spectra index maps lack the resolution to distinguish a flat-spectrum core feature. At 22 and 43\,GHz where the core might be dominant, the structure of 0954+556 is highly resolved on long baselines. Therefore, we disfavor identifying the emission seen at 22 and 43\,GHz as the core. The core might be intrinsically faint, blended with nearby jet emission, or self-absorbed even at higher frequencies.

\subsection{Intermittent activity}
The main morphology of 0954+556 in this study is consistent with previous work \citep{1995A&AS..110..213R,2005A&A...434..449R,2011ApJ...738..148M}, a triple structure at kpc scales, and a two-component structure located in the north--south direction at pc scales, which naturally raised a question of jet realignment. Regarding this concern, \citet{2005A&A...434..449R} proposed that 0954+556 could be a case of a smothered source or of recurrent activity, involving two distinct phases responsible for the kpc-scale and the pc-scale structures, respectively, as inferred from the total intensity structure. Our spectral analysis provides clear evidence that the observed emitting components in 0954+556 emerged from independent episodes of activity, rather than the pc-scale jet is now feeding the kpc-scale structure to form a continuous jet extending from pc to kpc.

0954+556 has experienced at least two episodes of jet activity, forming the triple structure at kpc scales and a two-component structure at pc scales. 
The two-sided jets are clearly detected at both spatial scales. The kpc-scale jets are collimated with a position angle of $\sim$-60$^\circ$. At pc scales, the high-resolution VLBA images revealed the north--south structure with a bending counter-jet and a bending forward jet.   
The estimated radiative ages of the outermost and innermost regions differ by over two orders of magnitude, approximately 0.3--1.4 $\mathrm{Myr}$ for component NW and 1--4 $\mathrm{kyr}$ for component S.
Three prominent components, i.e., component NE at the kpc scales, component NW at the kpc scales, and component S at the pc scales, are notably misaligned, and the counter-jets corresponding to components NW and S are detected, so it might lead us to assume that component NE represents a third jet activity. However, due to insufficient data, we were unable to perform a spectral fitting to derive a reliable radiative age for component NE.
Furthermore, from 1.4 to 5\,GHz, the two components NE and NW have similar spectral index distributions. At the 5--8.4 GHz spectral index map, although the component NE is slightly steeper than component NW, the opposite kpc-scale radio jets or lobes in double-lobed galaxies and quasars often have different spectral index distributions \citep{1991MNRAS.249..343L,1997MNRAS.289..753D}. Thus, we consider that components NE and NW originated from the same jet activity. The radiative age of the kpc-scale structure estimated by component NW reflects that the jet activity is on a short timescale, which allows us to detect the remaining emission from the earlier jet activity, i.e., the tail extending westward in VLBA images at 1.7 and 5\,GHz.

In addition, the low RM in 0954+556 proves the intermittent activity indirectly. Young radio galaxies such as CSOs are typically embedded in the dense interstellar medium of their host galaxies, exhibiting relatively high RM between $10^2$ and $10^4\,\mathrm{rad}\,\mathrm{m}^{-2}$ \citep{2010MNRAS.402...87A,2016AN....337...42R,2016MNRAS.459..820T} and low fractional polarization of less than 1$\%$\footnote{\url{https://science.nrao.edu/facilities/vla/docs/manuals/obsguide/modes/pol}} \citep{2021A&ARv..29....3O}. In contrast, the RM detected in 0954+556 is close to zero, and its $m$ is relatively high, increasing from 3.8$\%$ at 1.7\,GHz to 7.5$\%$ at 8.4\,GHz. The negligible RM indicates that the ambient medium surrounding the source is extremely tenuous, limiting strong low-frequency depolarization, and the high $m$ is therefore plausible. Such conditions can arise naturally if the source undergoes recurrent jet activity.
In the case that the re-collapse phase of the radio structure lasts longer than the repetition timescale, the repetitive shocks may maintain the interstellar medium warm and efficiently drive the gas out of the host galaxy. The process was described in detail by \citet{2009ApJ...698..840C} and \citet{2010ASPC..427..326S}. It is more likely that 0954+556 is a CSO maintained by intermittent jet activity rather than a single nascent outburst confined by a dense interstellar medium.

\subsection{Jet reorientation in intermittent activity}
0954+556 is a rare case in which the jet direction changed with different restarting activity. The earliest jet activity produced the forward linear jet and the corresponding counter-jet, which is 60 degrees north by west at kpc scales, and the discrete component NE, which is northeast of the core. The counter-jet and forward jet with a roughly north--south direction at pc scales were produced by the latest active phase. 
There are only a few reported cases of jet reorientation, all of which are related to a radio galaxy that harbors a blazar core. PBC J2333.9-2343 is the first one with significant jet reorientation ascribed to restarting activity, transitioning from a radio galaxy to a blazar \citep{2017A&A...603A.131H,2023MNRAS.525.2187H}, but no reasonable explanation has been proposed for the new jet pointing toward us by chance. Subsequently, \citet{2018ApJ...868...64P} revealed that the change in jet direction of B1646+499 could be explained by episodic jet activity with a precessing axis. In addition, a pilot study \citep{2022MNRAS.514.2122P} supplemented seven blazars characterized by exceptionally extended radio galaxy morphology several hundred kiloparsecs in size, and considered that a sudden jet realignment could be caused by a galaxy merger for 5BZU\,J1238+5325, by jet precession for 5BZU\,J1345+5332, and by the realignment of the SMBH spin as an interaction with the spin of the outer accretion disk for the remaining sources.

Two broad-line systems were not found in the Sloan Digital Sky Survey (SDSS) optical spectrum of 0954+556.\footnote{\url{https://skyserver.sdss.org/dr16/en/tools/explore/summary.aspx?ra=149.40910&dec=55.38271}} Nevertheless, this does not definitively exclude the possibility of a recent merger as a binary SMBH system could remain undetected if one SMBH lacks an emission-line system, if the emission lines from the two SMBHs are blended into a single, unusually broad profile, or if obscuration hides one system \citep{2010MNRAS.408.1103L}.
Meanwhile, the morphology of a non-blazar source is not affected strongly by jet precession since a small intrinsic bend is amplified by a small viewing angle along the line of sight.  
It was proposed that the intermittent jet activity on a timescale between $10^3$ and $10^6$ years is caused by a radiation pressure instability within an accretion disk \citep{2002ApJ...576..908J,2009ApJ...698..840C,2010ASPC..427..326S}.
Under this hypothesis, the instability occurs when the radiation pressure overtakes the gas pressure in the disk.
The duration of the activity and the outburst separation are determined by the mass of the central BH, the accretion rate, and the viscosity parameter.
The virial mass of the supermassive black hole ($M_\mathrm{BH,vir}$) in 0954+556 is $\log  (M_\mathrm{BH,vir}/M_\mathrm{\odot})=8.97\pm0.4$, and its Eddington ratio, which is the ratio of bolometric luminosity to Eddington luminosity, has been estimated to be $\log(L_{\mathrm{bol}}/L_{ \mathrm{Edd}})=-0.6$ by fitting the profile of the $\operatorname{Mg}_{I I}$ line \citep[The Sloan Digital Sky Survey Data Release 7]{2011ApJS..194...45S}.
In such a condition, if the viscosity parameter takes a value between 0.2 and 0.02, the activity will last for $10^3-10^4$ years and the separation time between episodes will be $10^{4.0}-10^{4.5}$ years \citep{2009ApJ...698..840C}. Considering that our analysis gives the lower limits, the implied timescales are consistent with our results in the order of magnitude.

\section{Conclusions}
\label{sec:sum}
We investigated the complex jet morphology of AGN 0954+556, uncovering evidence of intermittent jet activity. Our analysis, based on multifrequency data from the VLA and VLBA, has led to several significant findings.

1. We prefer to classify 0954+556 as a CSO rather than a FSRQ. In our results, 0954+556 shows weak variability in the radio and $\gamma$-ray bands, supported by Fermi-LAT data over 15 years, OVRO archival data over 2 years, and our multiepoch VLBA observations during the period from 2002 to 2016. 

2. The potential pc-scale radio core of 0954+556 is most likely near the position of the peak feature in the VLBA maps. This is in agreement with the optical centroid reported by \textit{Gaia}, the two-sided multifrequency jet structure, the spectral index distribution, and the spine-sheath polarization structure.    

3. We detected diffuse counter-jet components at kpc scales and at pc scales and a pc-to-kpc-scale bridge emission region for the first time. This confirms that the forward jet bends about 120 degrees from pc to kpc scales.

4. Together with the spectral index map and spectral aging analysis, we interpreted the apparent complex jet structure from pc to kpc scales as a consequence of at least two episodes of jet activity and rapid jet reorientation on timescales of one million years, which is possibly caused by accretion disk instabilities.

\section{Data availability} 
The reduced images associated with this work are available at the CDS via anonymous ftp to cdsarc.u-strasbg.fr (130.79.128.5) or via http://cdsweb.u-strasbg.fr/cgi-bin/qcat?J/A+A/.

\begin{acknowledgements}
We thank the anonymous referee for constructive comments that helped us improve the manuscript. 
This work supported by the National Key R\&D Program of China (2018YFA0404602). 
This work has made use of data from VLA and VLBA, both operated by The National Radio Astronomy Observatory, which is a facility of the National Science Foundation operated under a cooperative agreement by Associated Universities, Inc., as well as the light curve data from The Fermi Large Area Telescope (LAT) Light Curve Repository (LCR) \citep{2023ApJS..265...31A}, on behalf of NASA and the Fermi Large Area Telescope Collaboration.
This research has made use of \texttt{vlpy} \citep{2018ApJ...854...17L} which is a Python package used for VLBI data analysis and NASA/IPAC Extragalactic Database NED which is operated by the JPL, California Institute of Technology, under contract with the National Aeronautics and Space Administration.
\end{acknowledgements}

\bibliographystyle{aa} 
\bibliography{sample631}

\begin{appendix} 
\onecolumn
\section{Extra figures}
 \begin{figure*}[h!]
 \centering
 	\includegraphics[width=1\textwidth]{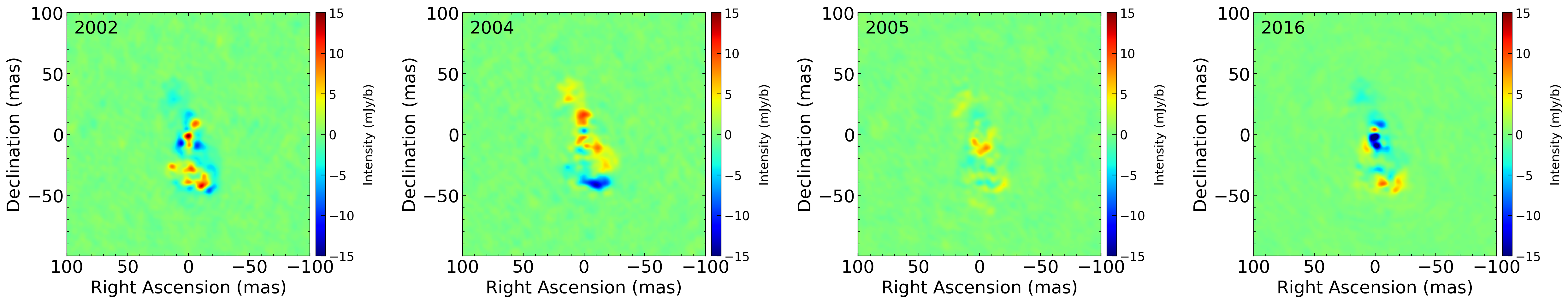}
     \caption{Subtraction maps to assess structural changes in 0954+556. The images in Figure~\ref{fig:5ghz-com} were convolved to a common restoring beam, corresponding to the beam from the observation on 13 June 2016. An average image was generated by taking the mean of the four epochs (2002, 2004, 2005, and 2016). Each individual image was then subtracted from the average image to highlight deviations. The color bar represents the intensity in units of mJy/b. The subtraction maps shows the intensity differences range approximately from -15\,mJy/b to 15\,mJy/b, indicating a stable structure of 0954+556.}
     \label{fig:subtract}
 \end{figure*}

\begin{figure*}[h!]
  \centering
 	\includegraphics[width=1.0\textwidth]{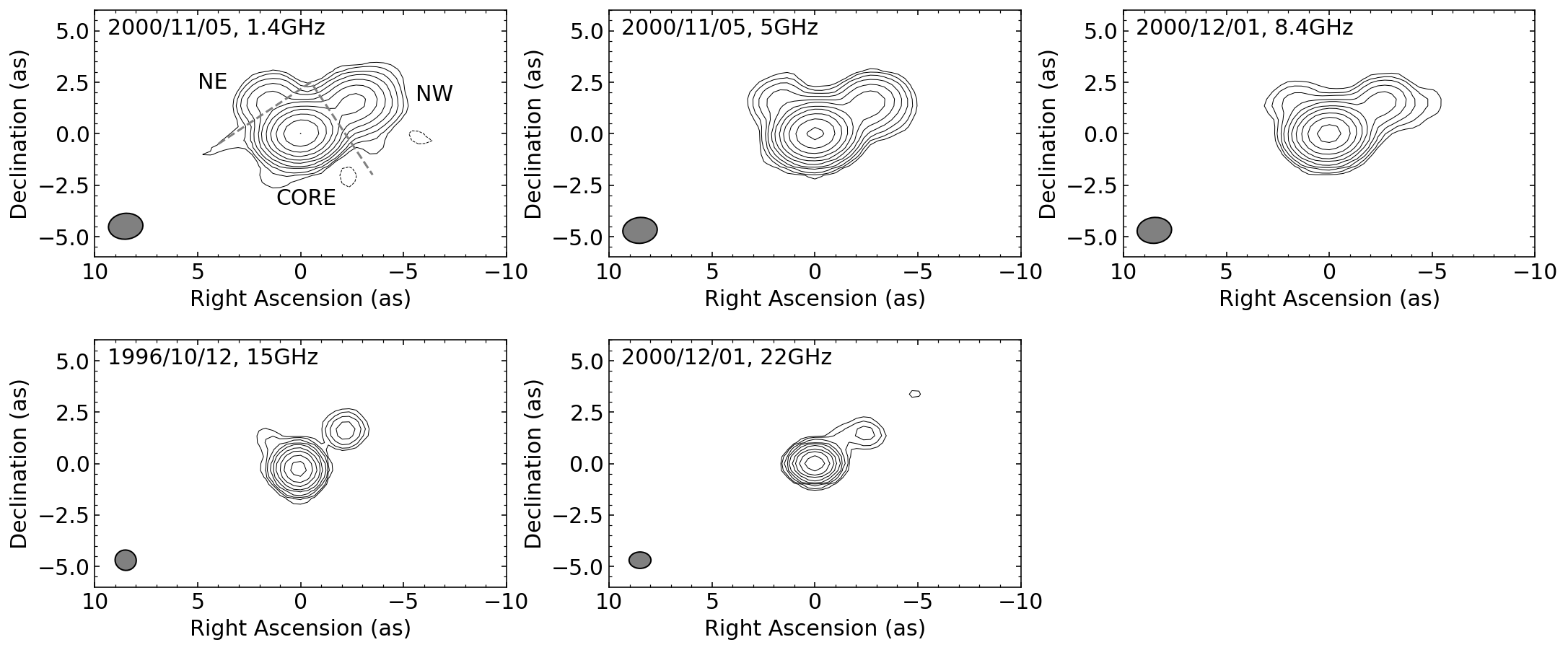}
     \caption{Images used when analyzing radiative age at kpc scale. The contours are drawn at -1, 1, 2, 4, 8, ..., of the first contour level (3$\sigma_\mathrm{rms}$). The synthesized beams are plotted in the bottom left corner of each image. Three components used in the spectral analysis, i.e., NE, core, and NW, are distinguished by two gray dashed lines.}
     \label{fig:rad-age-vla}
 \end{figure*}

 \begin{figure*}[h!]
 \centering
 	\includegraphics[width=1.0\textwidth]{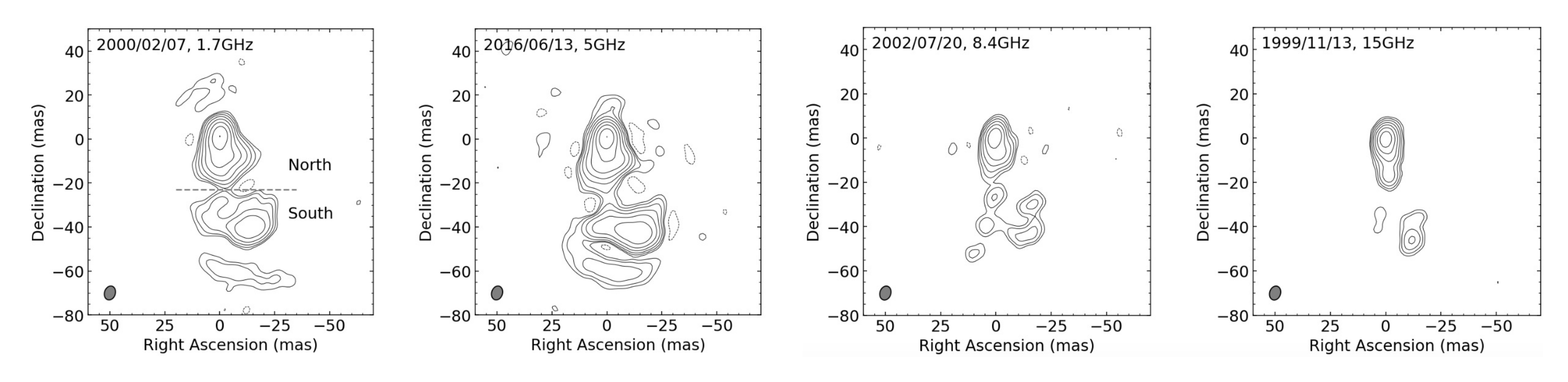}
     \caption{Images used when analyzing radiative age at pc scale. The contours are drawn at -1, 1, 2, 4, 8, ..., of the first contour level (3$\sigma_\mathrm{rms}$). The synthesized beams are plotted in the bottom left corner of each image. The division criteria for the north and south components are represented by a gray dashed line.}
     \label{fig:rad-age-vlba}
 \end{figure*}
\end{appendix}

\end{document}